\documentclass[pra,twocolumn,showpacs,aps,floatfix,amsmath,amssymb,amsfonts]{revtex4-1}

\usepackage{amsmath}
\usepackage{graphicx}
\usepackage{color}

\usepackage{mathrsfs}
\usepackage{amssymb,amsfonts,mathrsfs}
\usepackage{bm}

\newcommand{\beq}{\begin{equation}}
\newcommand{\eeq}{\end{equation}}

\newcommand{\x}{\mathbf{r}}
\newcommand{\bs}[1]{\boldsymbol{#1}}
\newcommand{\av}[1]{\left\langle #1 \right\rangle}

\definecolor{ao}{rgb}{0.0, 0.5, 0.0}

\begin{document}

\title{Magnetic gradiometer based on ultracold collisions}
\author{Tomasz Wasak$^1$}
\author{Krzysztof Jachymski$^{2}$}
\author{Tommaso Calarco$^3$}
\author{Antonio Negretti$^4$}
\affiliation{
$^1$ Faculty of Physics, University of Warsaw, Pasteura 5, 02-093 Warsaw, Poland\\
$^2$ Institute for Theoretical Physics III \& Center for Integrated Quantum Science and Technologies (IQST), University of Stuttgart, Pfaffenwaldring 57, 70550 Stuttgart, Germany\\
$^3$ Institute for Complex Quantum Systems \& Center for Integrated Quantum Science and Technologies (IQST), Universit\"at Ulm, 89069 Ulm, Germany\\
$^4$ Zentrum f\"ur Optische Quantentechnologien and The Hamburg Centre for Ultrafast Imaging, Universit\"at Hamburg, Luruper Chaussee 149, D-22761 Hamburg, Germany}

\date{\today}

\begin{abstract}
We present a detailed analysis of the usefulness of ultracold atomic collisions for sensing the strength of an external magnetic field as well as its spatial
gradient. The core idea of the sensor, which we recently proposed in K. Jachymski \emph{et al.}, Phys. Rev. Lett. {\bf 120}, 013401 (2018), is to probe the
transmission of the atoms through a set of quasi-one-dimensional waveguides that contain an impurity. Magnetic field-dependent interactions between the incoming atoms and
the impurity naturally lead to narrow resonances that can act as sensitive field probes since they strongly affect the transmission. 
We
illustrate our findings with concrete examples of experimental
relevance, demonstrating that for large atom fluences $N$ a sensitivity
of the order of 1 nT/$\sqrt{N}$ for the field strength and 100 nT/(mm $\sqrt{N}$)
for the gradient can be reached with our scheme.
\end{abstract}

\maketitle

\section{Introduction}

The precise detection of external fields is crucial for various technological applications as well as for basic science. For example, detecting electromagnetic and
gravitational fields is of paramount importance for time keeping and frequency standards~\cite{Bloom2014,Nicholson2015}, mineral discovery~\cite{Hoover1996},
navigation~\cite{DeGregoria2010}, medicine~\cite{Sander2012,Jensen2016}, material engineering~\cite{Balasubramanian2008}, and climate science~\cite{Chen2006}, but also
for precision measurements of fundamental constants~\cite{Webb1999,Chin2006,Zelevinsky2008,Blatt2008,Hudson2011,Baron2014}, for testing general
relativity~\cite{Schnabel2010,Schlippert2014,Biedermann2015,Rosi2017} or for seeking other effects foreseen by theories beyond the Standard Model~\cite{Ferrari2006,Salumbides2013,Borkowski2017}.

Improving the sensitivity of measurements can be accomplished in several ways. The most straightforward approach is to reduce the effect of noise sources with
technological improvements. However, in order to reach the fundamental precision limits dictated by quantum mechanics, it is
needed to optimize the initial quantum state of the system as well as the measurement process. This task is generally much harder to perform, but necessary to fully
exploit the quantum nature of a sensor~\cite{Giovannetti2004,Giovannetti2006}.

In the recent past, various strategies have been proposed to utilize cold atoms for quantum metrological purposes, including continuous probing of large atomic ensembles
with weak optical fields~\cite{Petersen2005,Petersen2006}, quantum non-demolition measurement and Kalman filtering
protocols~\cite{Chase2009a,Chase2009b,Negretti2013,Ciurana2017}, preparation of entangled (e.g., spin-squeezed) atomic
samples~\cite{Fernholz2008,Wasilewski2010,Riedel2010,Krischek2011}, Mach-Zender interferometry~\cite{Hardman2016}, and fountain clocks~\cite{Kruse2016}. 

In this work, we extend the ideas recently presented in Ref.~\cite{Jachymski2018}, where we proposed to exploit cold atomic collisions for high precision
magnetometry. Interestingly, compared to the previously discussed approaches, our detection scheme does not require either the preparation of entangled many-body states
or elaborated quantum measurement protocols and, importantly, it is robust against experimental imperfections (e.g., detector efficiency and finite temperature). Here, we
present the proposal in more details, extending our treatment beyond the $s$-wave interactions. We also show how the sensor can be used to estimate the spatial gradient of
the external field and provide a detailed derivation of the precision bounds.

\begin{figure}
\centering
\includegraphics[width=0.45\textwidth]{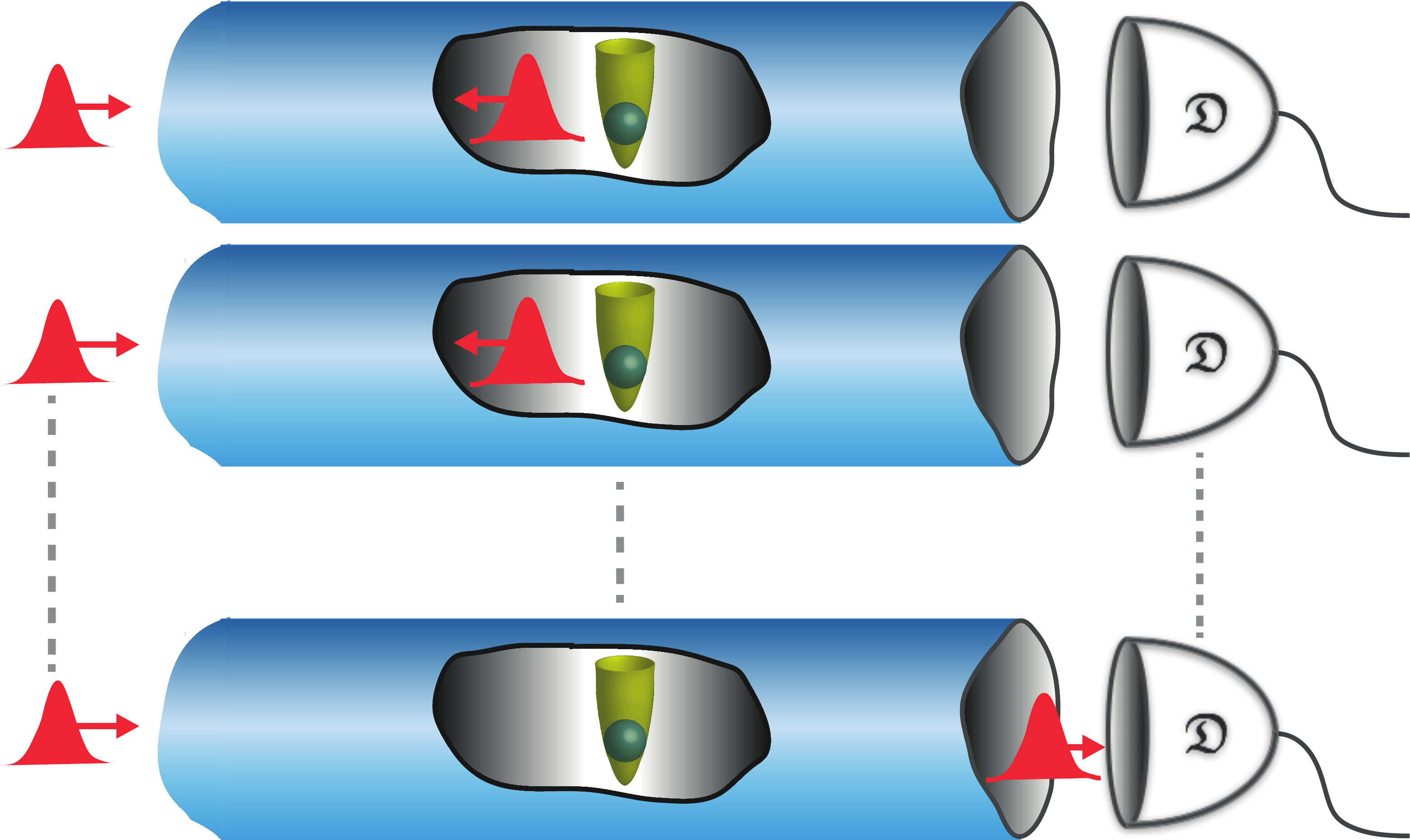}
\includegraphics[width=0.5\textwidth]{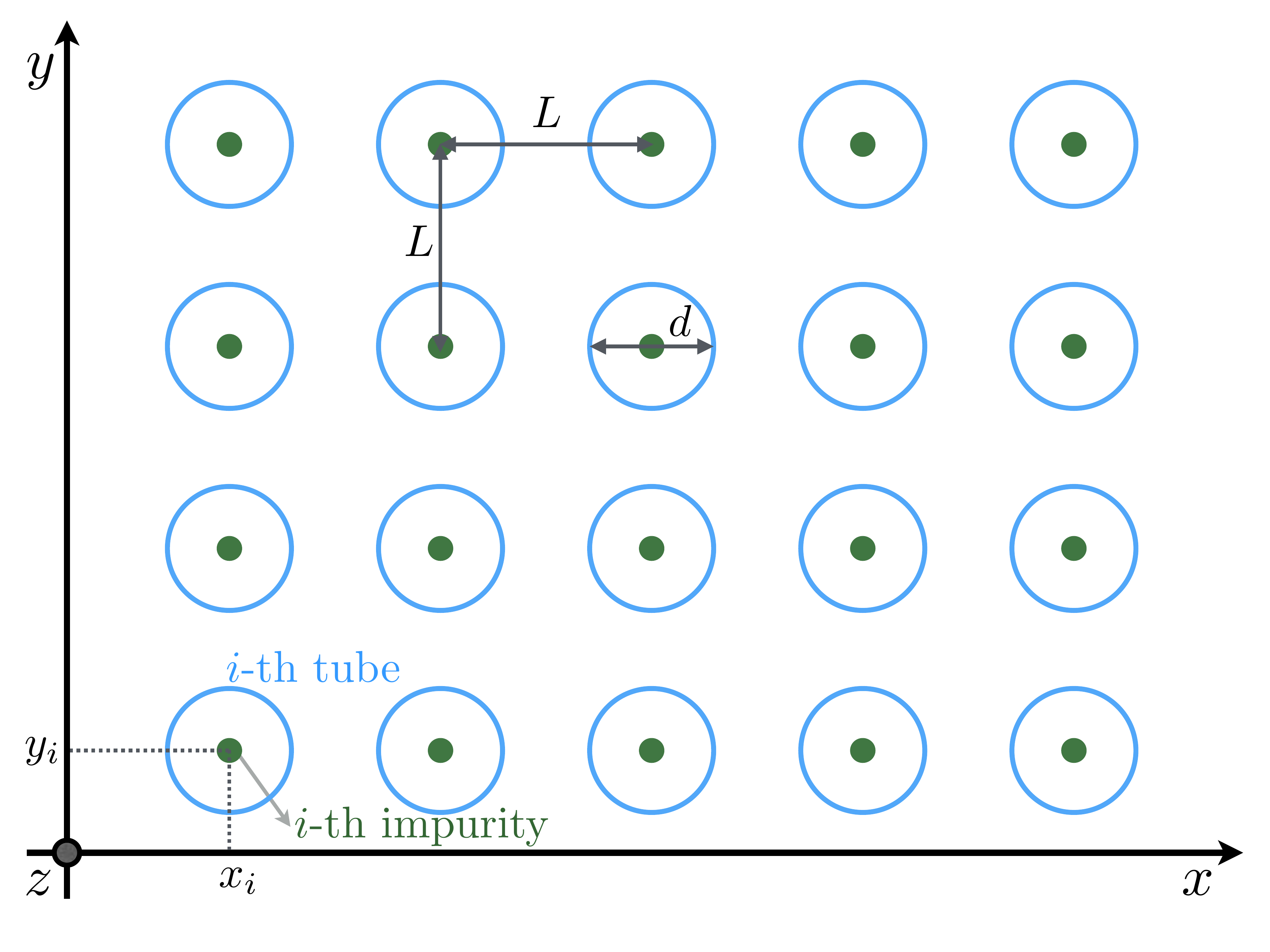}
\caption{\label{fig:setup} (Color online). Upper panel: Illustration of the magnetic field sensor that corresponds to the first column of blue circles in the plane outlined in
  the lower panel. $N$ atoms (red wave packets on the left of the upper panel) are sent to $N$ quasi-one-dimensional waveguides (blue cylinders or blue circles in the lower
  panel). A tightly confined impurity atom (green sphere) is placed in each waveguide. The transverse trap width $d$ is chosen close to the confinement-induced resonance
  condition (see text). The colliding atoms (red wave packets) can be either transmitted or reflected with probability depending on the external magnetic field
  strength. Transmitted atoms can then be detected outside the waveguide, e.g., by ionization and charge detectors ($\mathfrak{D}$).  Lower panel: Sketch of the section of the
  magnetic field sensor in the $x-y$ plane, whereas the $z$ axis is perpendicular to the plane and parallel to the symmetry axes of the waveguides (outwards with respect
  to the page). The (blue) circles indicate the waveguides of width $d$, whereas the full (green) circles the tightly trapped impurity atoms. The waveguides are
  separated from each other by a distance $L$.}
\end{figure}

For the sake of clarity, let us first briefly call the working principle of the collisional sensor, which is schematically shown in Fig.~\ref{fig:setup}. We consider an
ensemble of noninteracting atoms (red wave packets on the left-hand side of the upper panel). Each atom is injected into its corresponding quasi-one
dimensional waveguide (blue cylinders). Such a setup can be experimentally realized by means of a deep three-dimensional (3D) optical lattice with single-site access, which is then relaxed
in the longitudinal direction in a controlled way so that each atom acquires a longitudinal momentum (see, e.g. Refs.~\cite{Meinert2017,Robens2017}). In the centre
of each waveguide there is a tightly confined atom (green spheres), either of the same species or of a different one with respect to the moving atoms. The
collision can lead to transmission or reflection of the incoming atoms by the impurity. Transmitted (and possibly also reflected) atoms are then detected. The
sensitivity of the measurement on the magnetic field is due to a Feshbach resonance that controls the interaction strength between the atom and the impurity. It is
possible to tune the parameters in such a way that the probability of reflecting the colliding atom back from the impurity strongly depends on the local value of the magnetic
field.  The spatial spread of the waveguides (see lower panel in Fig.~\ref{fig:setup}) allows us to gather information about the average magnetic field strength at
their positions, therefore providing information about the field gradient.

The paper is structured as follows: In Sec.~II we describe atomic scattering in a quasi-1D waveguide in the vicinity of a Feshbach resonance. In Sec.~III we analyse the sensor performance by
providing experimentally relevant examples as well. Conclusions are drawn in Sec.~IV, whereas in the appendix we review briefly multiparameter estimation theory.

\section{Atomic scattering in a quasi-1D waveguide}

The problem of cold atomic scattering in quasi-one-dimensional confinement has been extensively studied in the
literature~\cite{Olshanii1998,Bergeman2003,Granger2004,Kim2006,Naidon2007,Giannakeas2012,Giannakeas2013,Peng2014,Hess2014,Idziaszek2015,Hess2015,Melezhik2016,Jachymski2017}. 
Here, we only provide a brief review of the most relevant results.

Assuming that the impurity atom is pinned in space (see Supplementary Material of Ref.~\cite{Jachymski2018} for a discussion of this assumption), the problem can be described
with the stationary Schr\"{o}dinger equation for the motion of the incoming confined atom [here ${\bf r}\equiv (x,y,z)$]:
\beq
\label{eq1}
\left[-\frac{\hbar^2}{2m}\nabla^2 + V_{\rm tr}(\mathbf{r}) + U(\mathbf{r})\right]\Psi(\mathbf{r})=E\Psi(\mathbf{r}).  
\eeq 
Here, $m$ is the mass of the atom, $V_{\rm tr}$ is the transverse harmonic trapping potential $V_{\rm{tr}}=\frac{1}{2}m \omega^2 \rho^2$ [with $\rho^2 = x^2 + y^2$ and
its characteristic length $d=\sqrt{\hbar/m\omega}$], and $U(\mathbf{r})$ is the potential resulting from the interaction with the impurity.

At large distances, the interaction vanishes and the wavefunction can be decomposed into the initial harmonic oscillator mode $\psi_{nm}$ in the transverse direction and
even and odd plane waves in the longitudinal direction.
In the general case, inelastic scattering can occur and lead to finite population of $n^\prime m^\prime$ modes. In the following, we restrict our considerations to the lowest mode of the transverse oscillator, which 
should provide the best conditions for precise
measurements. Extension of the calculations to higher modes is straightforward, albeit tedious.
The total energy of the incoming atom can then be given as a sum of the harmonic and unconfined part
\beq
E=\frac{\hbar^2 k^2}{2m}=\hbar\omega+\frac{\hbar^2 p^2}{2m}
\eeq
and is conserved during the collision, as the impurity is tightly trapped and cannot change its state. 
Note that we distinguish here the three-dimensional momentum $\hbar k$ from the one-dimensional $\hbar p$.

The scattering can be completely described in terms of two scattering amplitudes, 
which are related to the one-dimensional phase shifts $\eta_{\pm}$ by:

\beq
f^{(\pm)}(p)=-\frac{1}{1+i\cot\eta_\pm (p)}.
\eeq
In the case of identical bosons, symmetry allows only the even scattering. However, we are interested in the case of distinguishable particles, and thus we keep the odd
term for the sake of generality. The transmission coefficient, which describes the part of the flux that goes through the waveguide, can then be defined as~\cite{Olshanii1998}:

\beq \label{Tdef}
T(p)=\left|1+f^{(+)} + f^{(-)}\right|^2.
\eeq
This expression can be conveniently rewritten in terms of the phase shift as $T=\cos^2(\eta_++\eta_-)$.

We now need to connect the one-dimensional phase shifts to three-dimensional scattering quantities. This can be done analytically if the length scale characterizing
the interaction range is much smaller than the trap width $d$. One can then describe the scattering by a zero-range pseudopotential~\cite{Olshanii1998,Bergeman2003} or equivalently
use frame transformation techniques~\cite{Granger2004}. In general, even partial waves $\ell=0,2,\dots$ contribute only to the even part of the one-dimensional
scattering, while odd partial waves describe the odd part. Restriction to the $s$-wave ($\ell=0$) interaction results in~\cite{Olshanii1998}

\beq
\label{swave}
p\tan\eta_{+}(p)=-\frac{2}{d}\left(\frac{d}{a(k)}-\mathcal{C}\right)^{-1}.
\eeq
Here, $a(k)=-\frac{m}{\mu}\tan \delta_{\ell=0}(k)/k$ is the 3D energy-dependent scattering length that is rescaled by the $m/\mu$ factor due to our assumption of a pinned scattering center,
and $\mathcal{C}=-\zeta_H\!\left(\frac{1}{2},\frac{3}{2}-\frac{E}{2\hbar\omega}\right)$ with $\zeta_H$ being the Hurwitz zeta function.

Higher partial waves can be especially important for long-range interactions at low energies due to the different threshold laws.
For $p$-wave interactions, one obtains the following contribution to the odd phase shift~\cite{Granger2004}

\beq
\label{pwave}
\tan\eta_- \!=\! -\frac{6V_p(k) p d}{d^3} \left[1\!-\!12\frac{V_p(k)}{d^3}\zeta_H\!\left(-\frac{1}{2},\frac32-\frac{E}{2\hbar\omega}\right)\right]^{-1},
\eeq
where $V_p(k)=-\frac{m}{\mu}\tan\delta_{\ell=1}(k)/k^3$ is the 3D $p$-wave scattering volume.

Inclusion of the $d$-wave in the potential modifies the even part of the scattering as~\cite{Giannakeas2012}:

\beq
\label{dwave}
\begin{split}
p\tan\eta_{+}(p)=-\frac{1}{1+a(k) \mathcal{C}/d}\left(2\frac{a(k)}{d^2} +10\frac{a_d(k)^5}{d^6}\times\right.\\
\left. \times \frac{\left(1+(\mathcal{C}-\mathcal{C}_4 /2)a(k)/d\right)^2}{1+a(k) \mathcal{C}/d+a_d(k)^5/d^5 (\mathcal{C}_2+\mathcal{C}_3 a(k)/d)}\right).
\end{split}
\eeq
Here, $a_d(k)=\frac{m}{\mu}\tan\delta_{\ell=2}(k)/k$ is the 3D $d$-wave scattering length, and $\mathcal{C}_i$ are again given by Hurwitz zeta functions. Note that in all the above formulas the scattering lengths are calculated at finite $k$, corresponding to the total energy of the atom including the transverse confinement.

\begin{figure}
  \centering
  \includegraphics[width=0.45\textwidth]{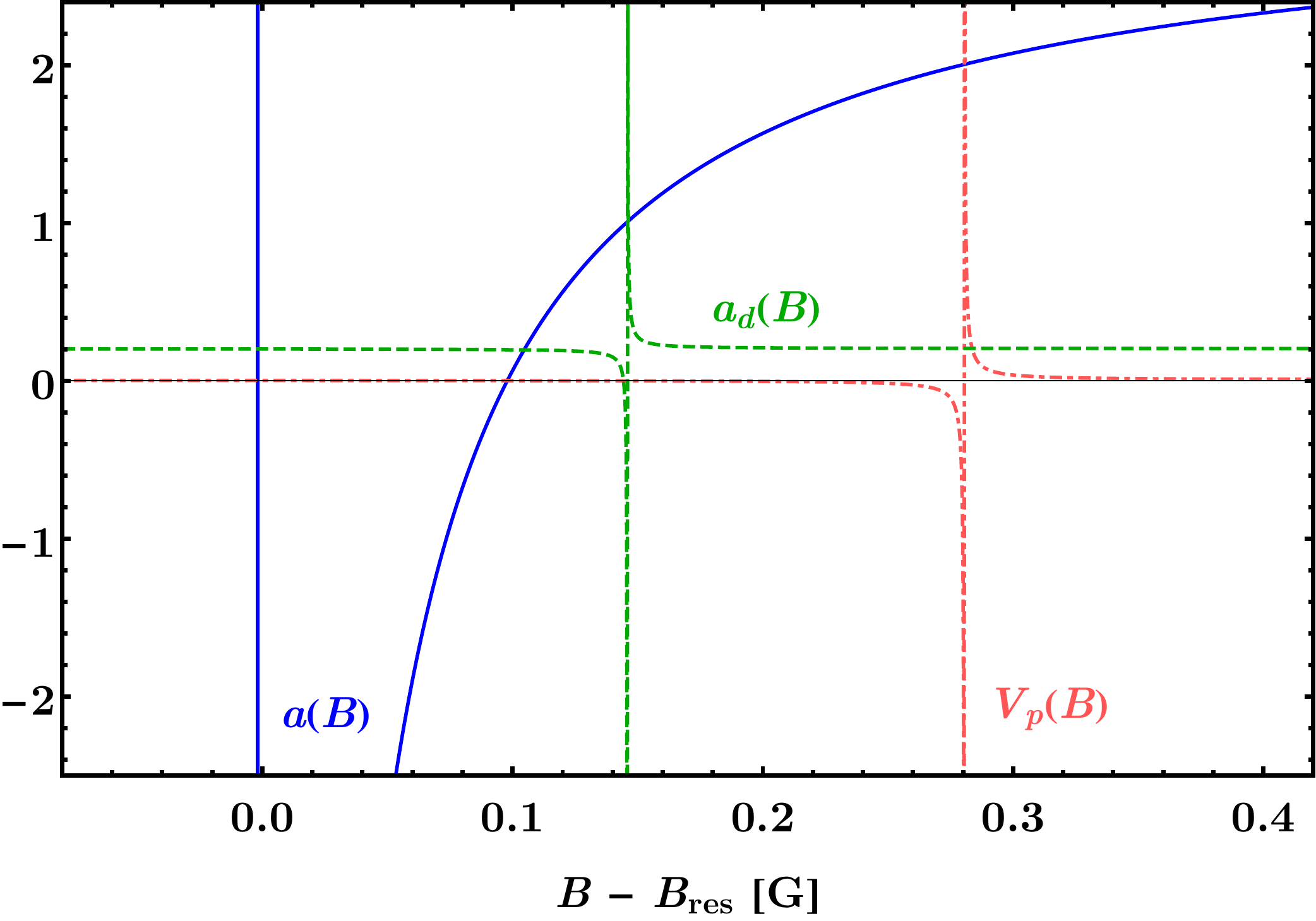}
  \caption{\label{fig:feshbach} (Color online) Scattering lengths in units of $\bar{a}$ computed for van der Waals scattering close to a Feshbach resonance. The blue solid
    line shows the $s$-wave scattering length, the red dash-dotted line the $p$-wave volume, and the green dashed line the $d$-wave scattering length.}
\end{figure}

Analytical formulas can be derived for arbitrary partial waves~\cite{Hess2015}. In general, higher partial waves lead to emergence of
additional very narrow resonances. These occur when their respective scattering lengths become comparable with the trap width, similarly to the $s$-wave case. In the
presence of a magnetic Feshbach resonance, all the scattering lengths corresponding to a partial wave can be tuned. 
The $s$-wave scattering length in the zero energy limit close to the Feshbach resonance is described by the universal formula~\cite{Chin2010}
\beq
a(B)=a_{\rm bg}\left(1-\frac{\Delta}{B-B_{\rm res}}\right)
\eeq
with $\Delta$ being the resonance width, $B_{\rm res}$ its position and $a_{\rm bg}$ the scattering length away from the resonance. Scattering in higher partial waves depends on
the details of the interaction potential. Here, we choose the interaction to have the van der Waals form $V(r)=-C_6/r^6$ with characteristic length 
$\bar{a}=2\pi (2\mu C_6/\hbar^2)^{1/4}/\Gamma(1/4)^2$, as defined in Ref.~\cite{Gribakin}, and $\Gamma$ being the Euler gamma function. This interaction is typical for the scattering of ultracold neutral atoms. We numerically solve the three-dimensional scattering problem with this potential in the presence of a Feshbach resonance and obtain the scattering phase shifts in
Eqs.~\eqref{swave}-\eqref{dwave} as a function of the magnetic field. We note that the analytic theory developed by Gao~\cite{Gao1998,Gao2000} predicts that
the $p$-wave scattering volume diverges, as a function of the magnetic field $B$, exactly at $a(B)=2\bar{a}$, and the $d$-wave scattering length at $a(B)=\bar{a}$.
Figure~\ref{fig:feshbach} 
shows the magnetic field dependence of the scattering lengths for an exemplary Feshbach resonance characterized by the width $\Delta=0.1$G. The higher partial wave
resonances occur exactly where expected.

\begin{figure}
  \centering
  \includegraphics[width=0.45\textwidth]{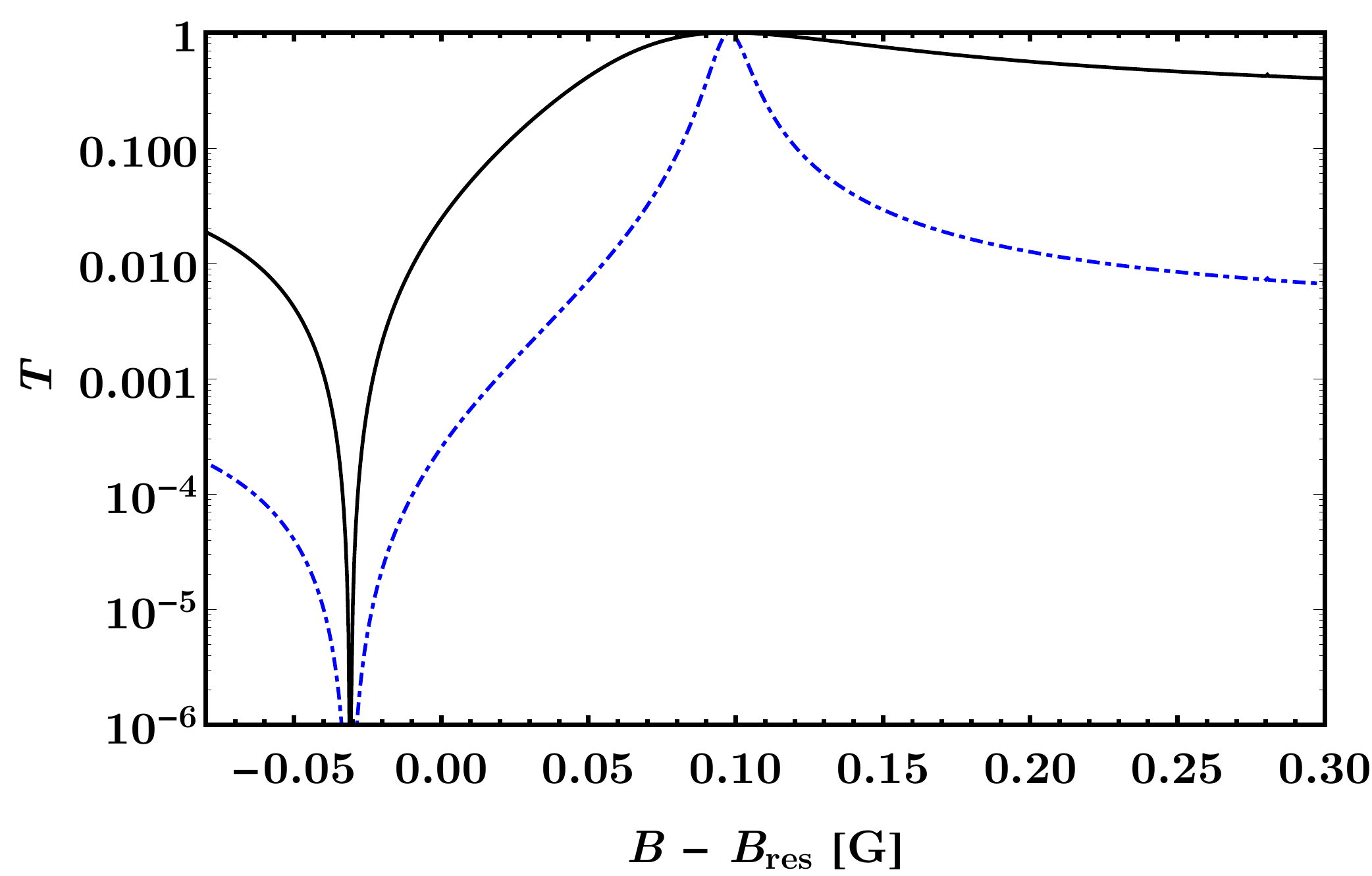}\\
  \includegraphics[width=0.23\textwidth]{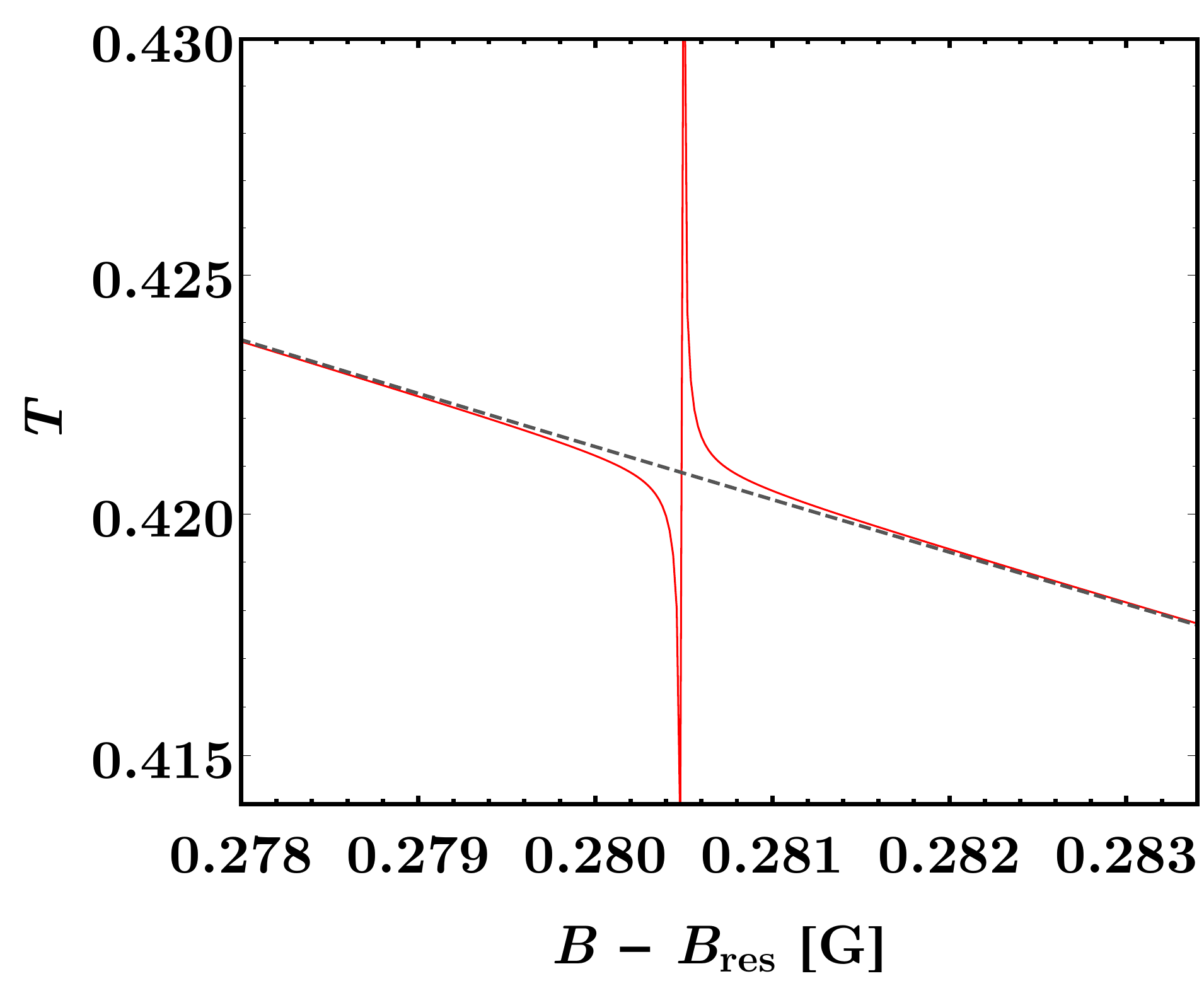}
  \includegraphics[width=0.23\textwidth]{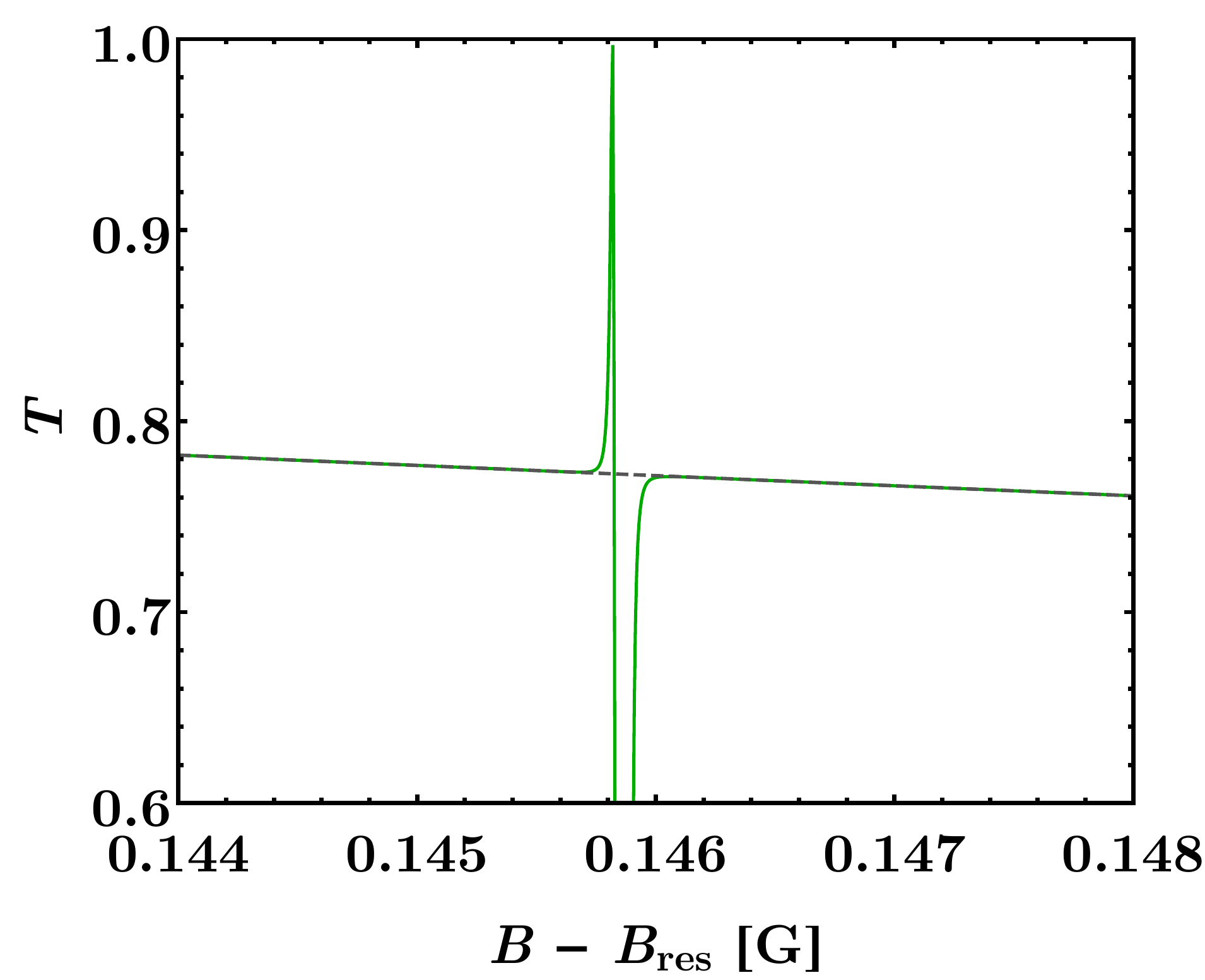}
  \caption{\label{fig:transm} (Color online) Transmission coefficient as a function of magnetic field calculated for the $s$-wave resonance presented in Fig.~\ref{fig:feshbach}
    assuming $d=20\bar{a}$ and $p=0.01\bar{a}^{-1}$ (black solid line) and $p=0.001\bar{a}^{-1}$ (blue dash-dotted line). The narrow resonances caused by higher
    partial wave scattering are not visible, but are shown in the lower panel for $p=0.01\bar{a}^{-1}$. The dashed (black) line gives the $s$-wave result, while the straight lines
    include higher partial waves, i.e. $p$-wave on the left (red solid line) and $d$-wave on the right (green solid line). A similar behavior occurs for lower values of $p$.}
\end{figure}

Having calculated the 3D scattering lengths, we can exploit Eqs.~\eqref{Tdef}--\eqref{dwave} to compute the transmission coefficient as a function of the magnetic field. The
results are presented in Fig.~\ref{fig:transm}. The insets show the narrow resonances resulting from the contribution of higher partial waves. Away from these resonances,
the transmission is well described by the simple model which includes only the $s$-wave scattering. 

While in the above calculation we assumed that the impurity is pinned in the center of the waveguide, this approximation can be relaxed. The motion of the impurity in a tight trap, even displaced from the center, can be included and will result in slight shift of the resonance positions as well as emergence of multiple narrow confinement induced resonances due to coupling of the center of mass and relative motion~\cite{Massignan2006}.

\section{Sensor performance}

The transmission of the atoms through the waveguides is strongly affected by the proximity of the scattering resonances. This phenomenon was used in
Ref.~\cite{Jachymski2018} to estimate the value of the magnetic field by exploiting a single waveguide. Below, we investigate the possibility of measuring the strength of
the magnetic field and, simultaneously, its spatial gradient utilizing an array of parallel waveguides (see also Fig.~\ref{fig:setup}).

To begin with, we will first perform a simple analysis of the precision achievable using an isolated single tube. 
Then, we apply the multiparameter estimation formalism discussed in the appendix to fully characterize the precision 
of the sensor for both the field value and of its gradient components.


\subsection{Field strength estimation with a single tube}
\label{sec:1tube}
  
In this scenario, the probability of detecting a transmitted or reflected atom is only sensitive to the strength of the magnetic field. 
In such a single-parameter estimation problem, the lower bound on the variance of an unknown
parameter is provided by the so-called classical Cram\'er-Rao theorem~\cite{braunstein1994statistical, refregier2012noise, cramer2016mathematical}.  Accordingly,
the ultimate attainable uncertainty of the estimated field at large atom numbers $N$ is given by
\beq 
\Delta B \geq \frac{1}{\sqrt N}
\frac{1}{\sqrt F},
\label{eq:prec}
\eeq 
where $N$ denotes the number of injected atoms into the tube (red wave packets in Fig.~\ref{fig:setup}). The scaling $N^{-1/2}$ is a statistical factor coming from the
increase of independent resources. 
The figure of merit of the sensor is given by the Fisher information $F$~\cite{refregier2012noise}
that is defined as
\begin{equation}
\label{cfi}
F\!= \! \sum_{s=\pm 1}\frac{1}{p(s|B)}\!\! \left(\!\frac{\partial p(s|B)}{\partial B}\!\right)^2.
\end{equation}
Here, $p(+1|B)\equiv T(B)$ is the  transmission probability, and $p(-1|B)\equiv 1-T(B)$ is the probability of reflecting the atom in the collision with the
impurity. The precision bound given by Eq.~\eqref{eq:prec} is saturated asymptotically by the maximum likelihood estimator in the limit of a large number
of atoms used in the estimation procedure. 

\begin{figure}
  \centering
  \includegraphics[width=0.46\textwidth]{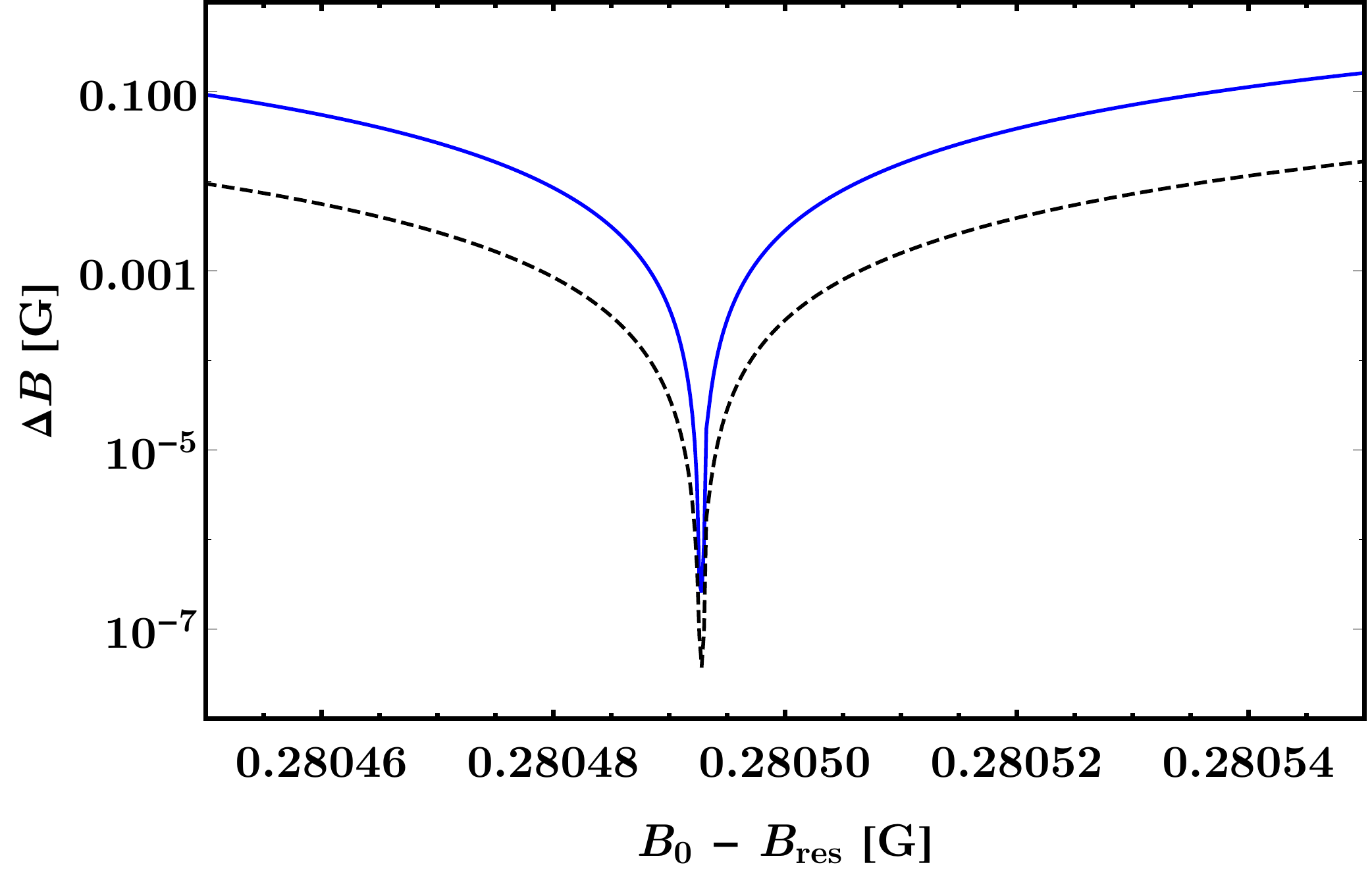}
  \caption{(Color online) Minimal uncertainty $\Delta B = F^{-1/2}$ [i.e. without $N^{-1/2}$, cf. Eq.~\eqref{eq:prec}] in units of~G for the narrow $p$-wave 
   confinement-induced resonance with $d=20 \bar{a}$ and $p=0.001\bar{a}^{-1}$ (solid blue line) and $p=0.01\bar{a}^{-1}$ (dashed black line).
  }\label{fig:prec}
\end{figure}

Expressing the probability distributions in terms of the transmission coefficient, $T(B)$, the Fisher information takes the following form:

\begin{equation}
  F = \frac{1}{T(B)[1-T(B)]} \left(\frac{d T(B)}{dB}\right)^2.
\label{fishers}
\end{equation}
The formula~(\ref{fishers}) implies that the lowest uncertainty is attained when the derivative of the transmission coefficient is the largest. The behaviour of $F$ in the vicinity of
$T=0$ and $T=1$ is determined by the dependence of $dT/dB$ close to these points. The case of the $s$-wave resonance was
analyzed in Ref.~\cite{Jachymski2018}. The uncertainty of the magnetic field $\Delta B$ in the vicinity of the $p$-wave confinement-induced resonance as a function of $B-B_{\rm res}$ is
displayed in Fig.~\ref{fig:prec} for $d=20 \bar{a}$ and two momenta: $p = 0.01 \bar a^{-1}$ (dashed black line) and $p = 0.001 \bar a^{-1}$ (solid blue line). One can notice that the
uncertainty for the larger momentum is smaller. This is intuitively explained by the fact the probability of detecting a particle without reflection is
increasing with momentum, but it rapidly drops to near zero value at the position of the confinement-induced resonance. The derivative of $T(B)$ is
larger if the change in transmission is bigger, which results in higher precision. With the $s$-wave interactions only, for the chosen parameters the achievable uncertainty is of the order of $10^{-3}$ G~\cite{Jachymski2018}, whereas at the $p$-wave
confinement-induced resonance we obtain $\Delta B \simeq 10^{-7}$ G. Such a big increase in precision is related to the fact that the $p$-wave resonance is much narrower than the $s$-wave. However,
the offset magnetic field has to be precisely controlled to ensure the atoms are close to the resonances in higher-partial waves. The results in Fig.~\ref{fig:prec} are improved further by a statistical factor of $1/\sqrt{M N}$, if $M$ waveguides are used with $N$ atoms per waveguide, as in Fig.~\ref{fig:setup}. 
Finally, we note that the above outlined observations and conclusions for the $p$-wave resonance apply to the $d$-wave resonance as well.

%
%

\subsection{Performance of the magnetic gradiometer}

We assume now that $M$ tubes are placed at fixed positions and that the magnetic field strength $B(\x)$ varies smoothly in space. 
In analogy to a single waveguide, 
we assign to the variable $\xi$ the value $+1$ if the atom is transmitted through the waveguide and
detected with (conditional) probability $p(\xi=+1|B)= T(B)$, and the value $\xi=-1$ if the atom is reflected after the collision with the impurity. 
For $M$ waveguides, we define a new random variable $\boldsymbol{\xi} = \{\xi_1,\ldots,\xi_M\}$. Thus, the probability of
transmitting or reflecting an atom is 
\begin{equation}
\label{joined}
  p(\boldsymbol{\xi}| B(\x_1),\ldots,B(\x_M)) = \prod_{i=1}^{M} p(\xi_i|B(\x_i)).
\end{equation}
Here $\x_i$ is the position of the impurity in the $i$-th tube, and $B(\x_i)$ is the corresponding magnetic field strength. Then, we 
expand the field strength up to
the first order, i.e. $B(\x_i) = B_0 + \nabla B \cdot \x_i$. Furthermore, we assume that all impurities are distributed on a plane, for which we set $z=0$ (see also
lower panel of Fig.~\ref{fig:setup}). Hence, we have
\begin{equation}
  B(\x_i) = B_0 + B_x x_i + B_y y_i,
\end{equation}
where $B_x\equiv \partial B/\partial x$ and $B_y\equiv\partial B/\partial y$ denote the $x$ and $y$ magnetic field gradient components, respectively, while $x_i$ and
$y_i$ are the corresponding coordinates in the plane. Thus, the parameters that we aim at estimating from the measurement records of the atom transmission and reflection
are the magnetic field strength $B_0$, and its gradient components $B_x$ and $B_y$. 

%
%

In order to apply the general formalism described in the appendix, we rename the three ($n=3$) unknown parameters as follows: $\gamma_0\equiv B_0$, $\gamma_1\equiv B_x$, and
$\gamma_2\equiv B_y$. Because of the additivity of the Fisher information matrix with respect to independent events, 
the FIM can be rewritten as a sum of Fisher information matrices describing each
waveguide separately: 
\begin{eqnarray}
\label{FIMsum}
  \mathbf{F} = \sum_{i=1}^M \mathbf{F}^{(i)}.
\end{eqnarray}
By denoting $T_i \equiv T(B(\x_i)) = T(B_0 + B_x x_i + B_y y_i)$ and its derivative with respect to $B$ by $T_i' \equiv T'(B_i) = T'(B_0 + B_x x_i + B_y y_i)$, the $i$-th
FIM takes the form of a product of a factor depending on the transmission coefficient by a matrix describing the geometry of the problem, i.e., depending only on the
positions of the waveguides:
\begin{equation}
\label{FIMi}
\mathbf{F}^{(i)}  = \frac{(T_i')^2}{T_i(1-T_i)}
\begin{pmatrix}
1 & x_i & y_i  \\
x_i & x_i^2 & x_i y_i  \\
y_i & x_iy_i & y_i^2
\end{pmatrix}.
\end{equation}
Hence, by performing the sum from Eq.~(\ref{FIMsum}) with the above outlined expression for $\mathbf{F}^{(i)}$, we obtain the full Fisher information matrix for $M$
tubes, which, according to Eq.~\eqref{crlb}, provides us the minimal attainable uncertainty for each of the three unknown parameters $B_0$, $B_x$ and $B_y$.  Let us
remark that at least three non planar tubes are necessary in order to obtain a meaningful estimation of the gradient in two spatial directions. Indeed,
mathematically, the structure of the matrix $\mathbf{F}^{(i)}$ enforces the determinant of $\mathbf{F}$ in Eq.~\eqref{FIMsum} to vanish unless $M > 2$. Moreover, we note
that the matrix $\mathbf{F}$ is not invertible for $M=3$ planar tubes.

In order to analyze the performance of the sensor, we consider exemplarily the case of $M=51\times 51$ tubes with equal spacing $L= 523$ nm in each direction (see lower panel
in Fig.~\ref{fig:setup}). For the sake of numerical simplicity, we assume that the field can change only along the $x$ direction, i.e., $B_y\equiv 0$. 
Notwithstanding, the additional $M$ rows of the tubes array (see lower panel in Fig.~\ref{fig:setup}) contribute
statistically due to the accumulated data.  Hence, in order to determine the minimal uncertainty, expressed by Eq.~\eqref{crlb}, of the magnetic field strength and its
spatial derivative along the $x$-axis, 
we can safely neglect in 
Eq.~\eqref{FIMi} the third column and the
third row.

\begin{figure}
\centering
\includegraphics[width=0.45\textwidth]{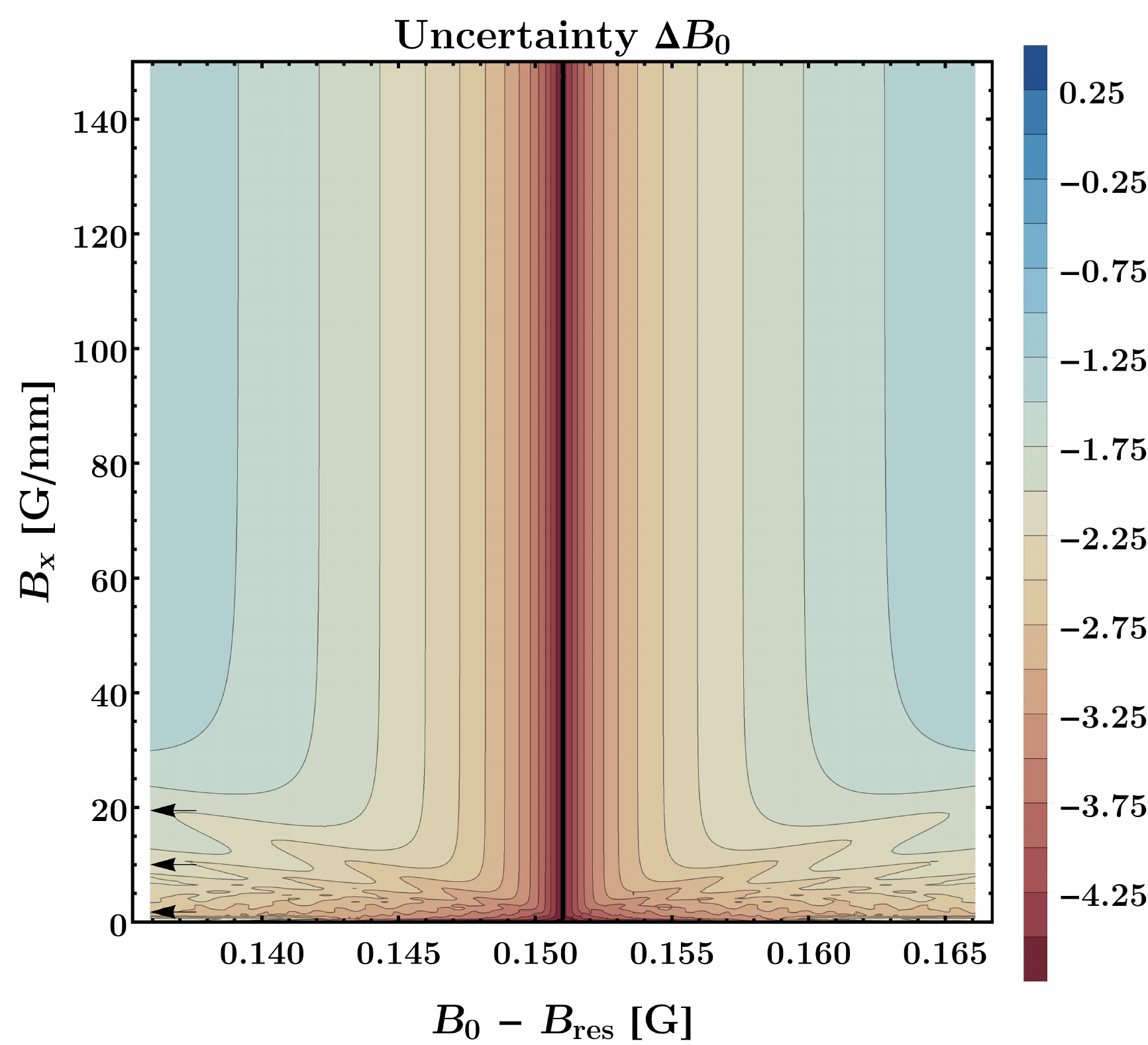}\\
\includegraphics[width=0.45\textwidth]{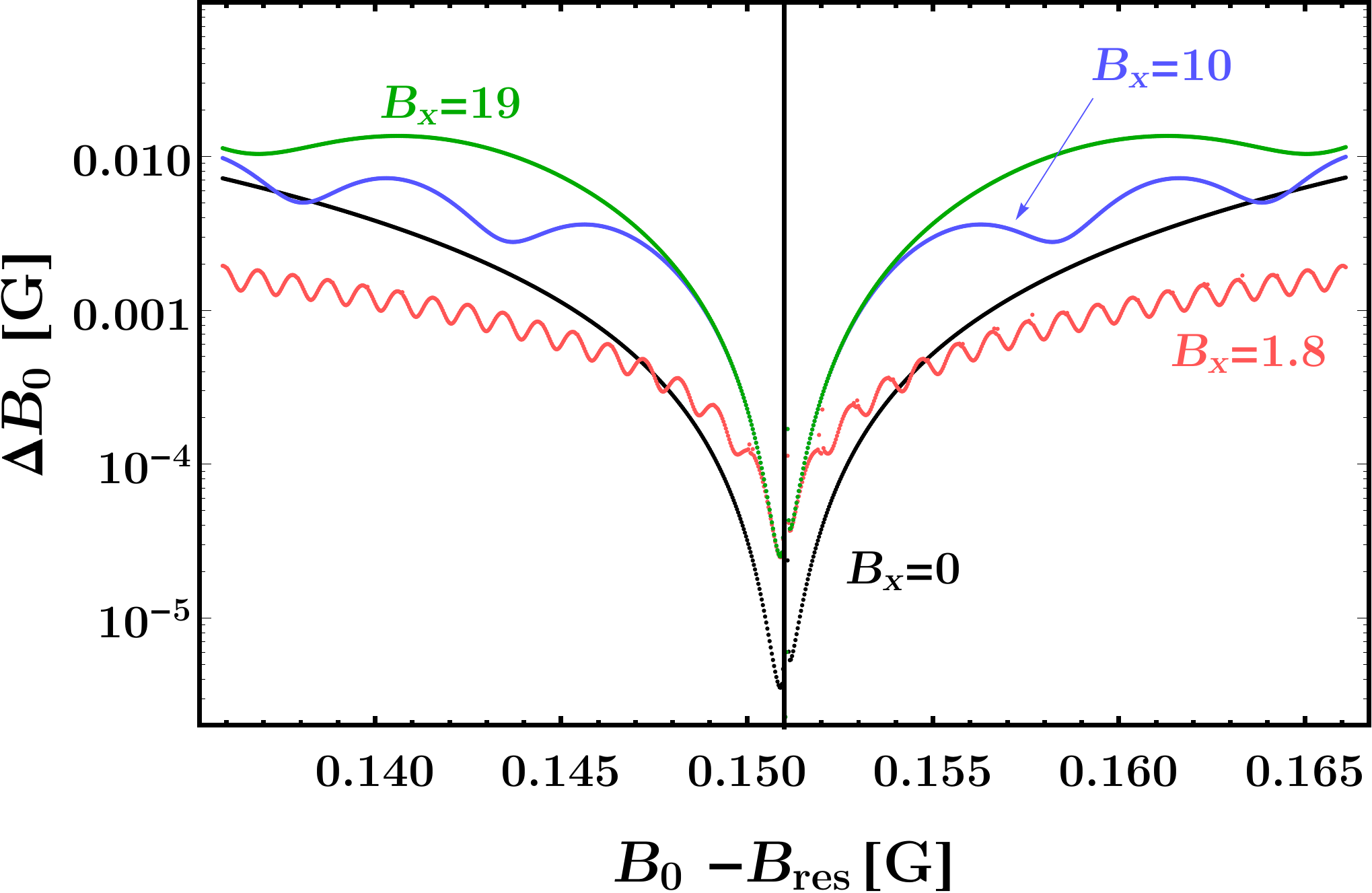}
\caption{(Color online)
  Upper panel: Estimation uncertainty $\Delta B_0$ of the magnetic field strength as a function of $B_0$ and $B_x$.  
  The color code (blue -- large $\Delta B_0$, red -- small $\Delta B_0$) is in logarithmic scale. 
  The sensor's best operating region is for $B_0-B_\mathrm{res} \approx \Delta$ (i.e., the resonance width), namely when the transmission is of the order of unity, and for small gradients. 
  Lower panel: Estimation uncertainty $\Delta B_0$ as a function of $B_0-B_\mathrm{res}$ for fixed gradient $B_x$. 
  The values of $B_x$ are indicated on the figure as well as with black arrows in the upper panel (bottom-left corner). 
  The black vertical line shows $B_0-B_\mathrm{res}=\Delta$.
}\label{fig:g0}
\end{figure}

In Fig.~\ref{fig:g0} we display the attainable uncertainty $\Delta B_0$ in the estimation of the magnetic field strengh $B_0$ measured with respect to the resonance position
$B_\mathrm{res}$. The parameters used in the calculation were $\Delta=0.15$G, $d=20\bar{a}$, $a_{\rm bg}=9.76\bar{a}$ (similar to caesium atoms) and $p=10^{-4}\bar{a}^{-1}$. For such a small
momentum the optimal working point is expected near the unit transmission region $B\approx\Delta$, in contrast to the higher energy case, where the zero transmission
region is favorable~\cite{Jachymski2018}.  As it can be seen from the figure, the uncertainty $\Delta B_0$ depends on both parameters $B_0$ and $B_x$.  For vanishing
gradient, the uncertainty changes appreciably around the optimal operating point $B_0\approx\Delta$, where $\Delta$ is the resonance width, at which the transmission
rapidly approaches unity (see also Fig.~\ref{fig:transm}).  At that point, the attainable uncertainty is on the order of $10^{-5}$ G, which is further enhanced by
the statistical factor $1/\sqrt{N}$ related to the number of atoms used in the protocol.  For the values of the field $B_0$ for which the transmission drops off by two orders
of magnitude, the uncertainty deteriorates as well and reaches at best the order of $10^{-2}$~G.

\begin{figure}[htb]
\centering
\includegraphics[width=0.45\textwidth]{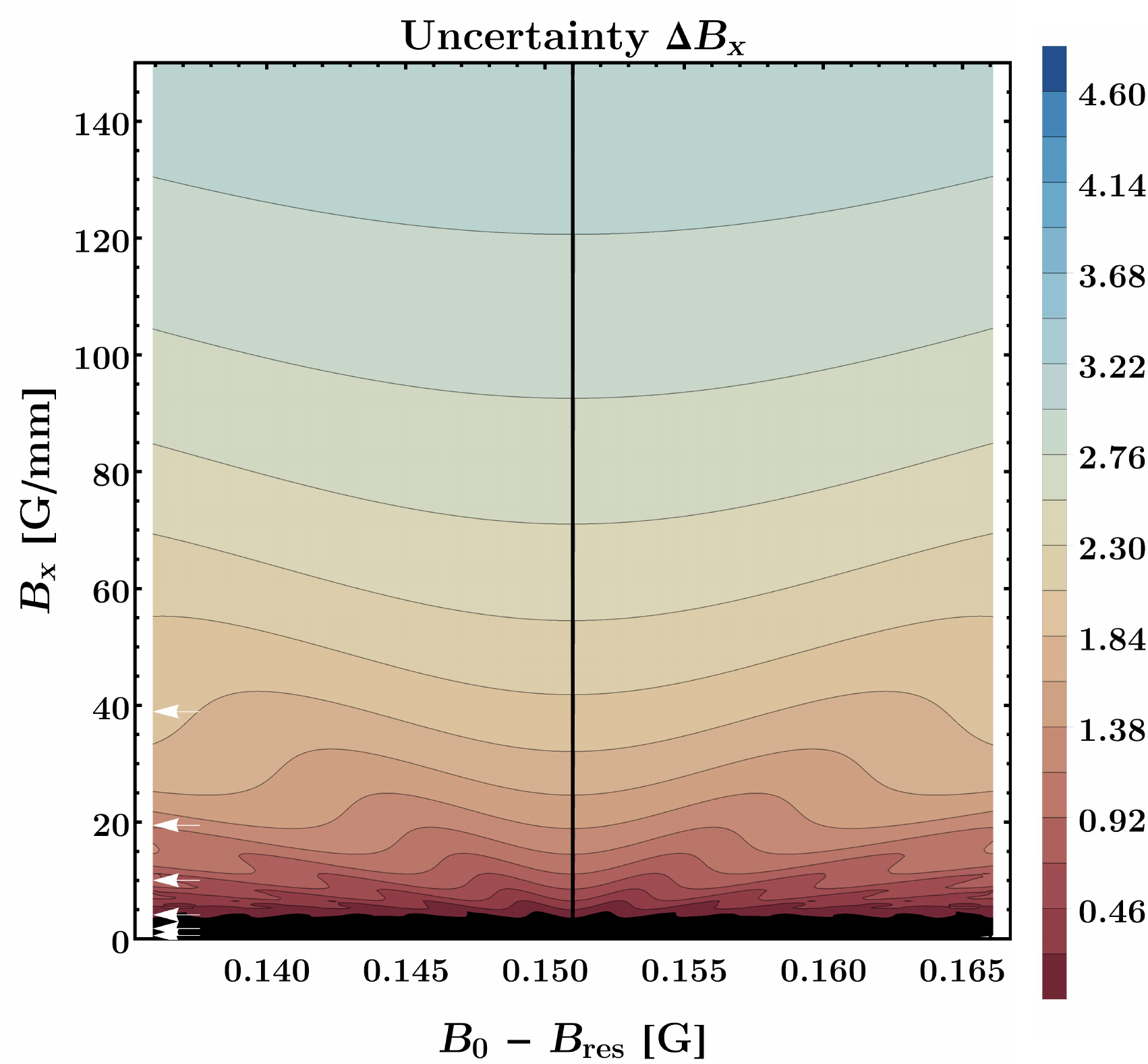}\\
\includegraphics[width=0.45\textwidth]{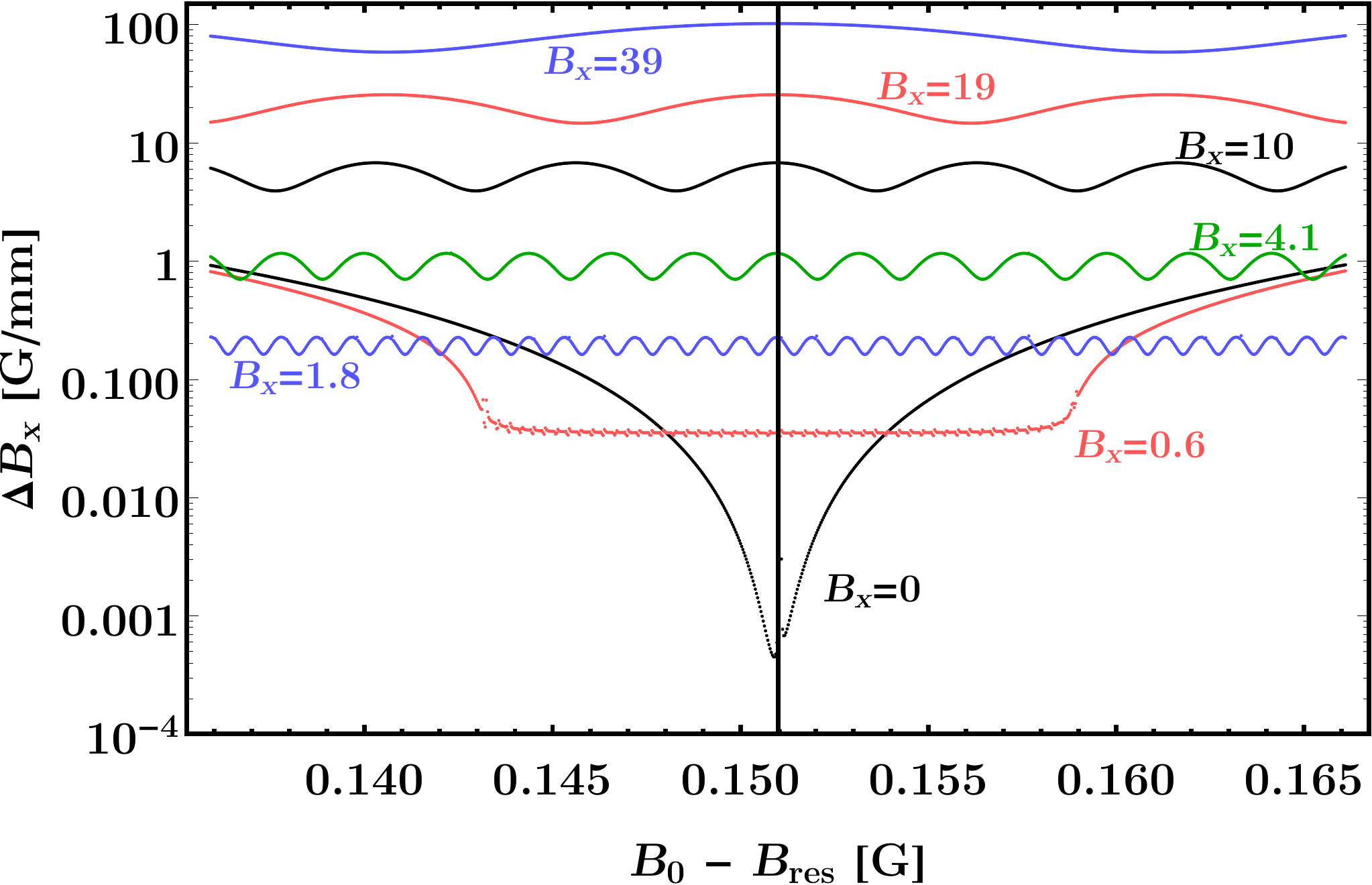}
\caption{(Color online)
  Upper panel: Estimation uncertainty $\Delta B_x$ of the magnetic field gradient as a function of $B_0-B_\mathrm{res}$ and $B_x$. 
  The color code (blue -- large $\Delta B_x$, red -- small $\Delta B_x$) is in logarithmic scale. 
  The sensor's best operating region is for $B_0-B_\mathrm{res}\approx \Delta$ (i.e., the resonance width),  when the transmission is of the order of unity, as well as for small gradients. 
  Lower panel: Gradient uncertainty $\Delta B_x$ as a function of $B_0-B_\mathrm{res}$ for fixed $B_x$. 
  The values of $B_x$ are indicated on the figure as well as with white arrows on the upper panel (bottom-left corner). 
  The black vertical line presents $B_0-B_\mathrm{res}=\Delta$. 
}\label{fig:gx}
\end{figure}

On the other hand, for a non-vanishing gradient, the uncertainty $\Delta B_0$ decreases by an order of magnitude, but it is maintained at the level of $10^{-4}$ G for 
fields around $B_0\approx\Delta$ and gradients up to few tens of Gauss per millimetre. Departure from $B_0\approx\Delta$ for any value of the gradient leads to the
increase of the uncertainty $\Delta B_0$.  In addition to this, as it can be seen from the Fig.~\ref{fig:g0} (lower panel), the precision $\Delta B_0$ exhibits
wavy features as a function of $B_0$ when departing from the region near to $B_0\approx\Delta$.  Furthermore, the ``frequency'' of this oscillatory behaviour is
decreasing as the gradient increases. This phenomenon can be understood as follows: Let us first suppose that the gradient is zero and the sensor is working around its
optimal operating point, namely $B_0\approx\Delta$.  In this case, all the terms of the sum in Eq.~\eqref{FIMsum} are of the same order and contribute a small
uncertainty.
However, when the gradient is small but non-zero, the local field at some of the waveguides is far from their optimal points, which decreases some terms in
Eq.~\eqref{FIMsum}, and, consequently, the uncertainty of estimation grows. As the field strength $B_0$ is varied, the local field at some of the waveguides approaches the optimal points, while at other waveguides the local field is away from them.
As a consequence, with the change of $B_0$, the uncertainty $\Delta B_0$ is in general higher than the one at the optimal point, but it exhibits periodic increases and
decreases, i.e. revivals. The same reasoning applies to $\Delta B_x$ (see Fig.~\ref{fig:gx}).

As it can be seen from Figs.~\ref{fig:g0} and~\ref{fig:gx}, the ``period" of such wavy features is larger for larger gradients. 
This can be understood in the following way:
For small gradients, the local field at the waveguides is very close to their optimal points, and thus the period should be small.
Instead, for large gradients, the local field is almost in all waveguides away from the optimal point and thus their contribution to the sum in Eq.~\eqref{FIMsum} is small, implying a small uncertainty.
Hence, the local field at a small number of waveguides will be near the optimal operating points as the field $B_0$ is varied, implying a larger period.


Finally, in Fig.~\ref{fig:gx}, we present the minimal attainable uncertainty $\Delta B_x$ for the estimation of the magnetic field gradient along the $x$ direction.
Similar to $\Delta B_0$, the optimal operating point is achieved when $B_0\approx\Delta$ and $B_x=0$. However, contrary to the estimation of the field, departure from
$B_x=0$ leads to an increase of the gradient uncertainty. Therefore, in this case, the performance of the sensor is the highest only for small gradients.  For the
parameters and geometry we have chosen, the uncertainty $\Delta B_x$ is the smallest and reaches the value $10^{-3}\ \mathrm{G\, mm}^{-1}$ for $B_0\approx\Delta$ and
small gradients.  For increasing gradients, the precision deteriorates and for $B_x \approx 1 \mathrm{G\, mm}^{-1}$ it is on the order of a Gauss per millimetre.
We remind, however, that the uncertainties $\Delta B_0$ or $\Delta B_x$ presented so far are without the statistical factor $1/\sqrt{N}$, which is inherent in the
statistical post-processing of the data and is related to finite resources used during the estimation procedure. By multiplying those uncertainties with that numerical
factor, we can further improve the sensitivity of the proposed sensor. Thus, also for increasing gradients, repeated measurements can improve the performance of the proposed gradiometer also away from the optimal operating point. 
Finally, similarly to $\Delta B_0$ and for the same reason discussed above, we observe that the precision $\Delta B_x$ exhibits an oscillatory behaviour as a function of the magnetic
field $B_0$ (see Fig.~\ref{fig:gx} lower panel).   

\section{Summary and conclusions}

We have proposed a new magnetic field sensor scheme utilizing atomic collisions in an array of waveguides.
At the input of the device, single atoms are injected into the waveguides, and then collide with the
impurities, while the transmitted and reflected atoms are detected at the end of each waveguide. From such a measurement recording, we infer the values of
the characteristic properties of the external magnetic field, that is, the strength of the field and its gradient along two directions.


We 
provided the attainable values of the uncertainties of the field characteristics.  In our previous
work~\cite{Jachymski2018}, we proposed a sensor operating only on a single waveguide and we showed that it is possible to reach an uncertainty of the field strength of the
order of nT/$\sqrt{N}$. 
Here, we extended the concept and showed that the multi-waveguide configuration of the sensor can be exploited to simultaneously measure magnetic field strength and its gradient with a precision on the order of 1 nT/$\sqrt{N}$ and 100 nT/(mm $\sqrt{N}$), respectively. 

The achievable precision can be still improved by a number of strategies. First, the simplest way to reduce the uncertainty is by increasing the number of
atoms at the input, which improves the statistical scaling of the uncertainty. Second, the possibility to tune the system 
(e.g., by controlling the frequency of the transverse trap) 
into the vicinity of the scattering resonances in the higher partial waves can decrease the uncertainty of the magnetic field by a
few orders of magnitude. Such an approach, however, is experimentally demanding since the offset field has to be precisely controlled.
We note that long-range interactions between atoms and the impurity will also generate additional narrow resonances, which might be employed for metrology leading 
to a similar sensor's performance as in the case of resonances in higher partial waves.
Lastly, the use of initial
entangled states can in principle lead to a further decrease of the uncertainty due to quantum correlations in a similar manner as squeezed or GHZ states employed in photonic quantum metrology~\cite{Giovannetti2006}. For this last strategy, however, one needs to devise
experimentally feasible protocols for engineering entangled input states and collective measurements.

To conclude, our work demonstrates that ultracold atomic collisions are useful for quantum sensing. Since Feshbach resonances can be controlled by many techniques, the
device we propose can find various applications in ultracold laboratories, where the magnetic field and its spatial characteristics have to be precisely known, e.g.,
in quantum simulators based on atoms in optical lattices~\cite{Gross2017}. In the current state, the sensor operates in a limited magnetic field range close to the Feshbach resonance and as such requires calibration with respect to the resonance position. It would be desirable to get rid of this requirement, possibly providing a more universal metrological standard in the future.

\begin{acknowledgments}
We are grateful to Paul S. Julienne and Zbigniew Idziaszek for valuable discussions. This work was supported by the Alexander von Humboldt Foundation, the Polish National Science Center Project No. 2014/14/M/ST2/00015, the Cluster of Excellence The Hamburg Centre for Ultrafast Imaging of the Deutsche Forschungsgemeinschaft, and the European Union FP7 FET Proactive
Project DIADEMS (Grant No. 611143).
\end{acknowledgments}

\appendix

%
%

\section{Multiparameter estimation formalism}
\label{eq:gen-est}

Here we describe the general formalism used in the main text in order to make use of the multiparameter estimation~\cite{helstrom1976quantum}. 

To begin with, let us denote the vector containing $n$ unknown parameters to be estimated by $\bs{\gamma}=(\gamma_0,\gamma_1,\ldots,\gamma_{n-1})$.  Specifically to our problem, 
the parameter $\gamma_0$ is the strength of the magnetic field at some arbitrary point in space, whereas $\gamma_1$ and $\gamma_2$ are
the $x$ and $y$ components of the field gradient (cf. Fig.~\ref{fig:setup}); this is an estimation of $n=3$ unknown parameters.

The parameters $\bs{\gamma}$ have to be extracted from the measurement record of a certain experimentally accessible observable $\bs{\xi}=(\xi_1,\xi_2,\ldots,\xi_M)$. In our setting, $\bs{\xi}$ denotes a reading of the detectors situated at the end of each of the $M$ waveguides, yielding the number of transmitted atoms. More specifically, if an
atom is recorded at the $m$-th waveguide, the component $\xi_m$ of $\bs{\xi}$ is assigned a prescribed value, taken arbitrarily equal to, say, $+1$.

After $N$ of such independent experimental runs, a data set $\bs{\xi}^{N} = \{\bs{\xi}^{(1)}, \bs{\xi}^{(2)}, \ldots, \bs{\xi}^{(N)}\}$ is obtained, where
$\bs{\xi}^{(i)}=(\xi_1^{(i)},\xi_2^{(i)},\ldots,\xi_M^{(i)})$ denotes an outcome of the $i$-th experimental run of the observable $\bs{\xi}$.  The probability of
obtaining $\bs{\xi}^{(i)}$ in an experimental run is given by $p(\bs{\xi}^{(i)}|\bs{\gamma})$, i.e., it is conditioned on the actual value, yet unknown, of the parameters
to be determined.  Thus, the outcomes $\bs{\xi}^N$ of the $N$ runs of the experiment, assuming statistical independence, are governed by a joint probability distribution
$P(\bs{\xi}^N) = p(\bs{\xi}^{(1)}|\bs{\gamma})p(\bs{\xi}^{(2)}|\bs{\gamma})\cdots p(\bs{\xi}^{(N)}|\bs{\gamma})$.  Hence, the distribution of the observed outcomes
$\bs{\xi}^N$ is governed by the underlying values of the parameters $\bs{\gamma}$, too.  The inference about the value of the unknown parameter vector $\bs{\gamma}$ is
drawn from the data $\bs{\xi}^N$ by means of a certain function $\bs{\Gamma}$ of the acquired measurement data. 
The function $\bs{\Gamma}(\bs{\xi}^N) = (\Gamma_0(\bs{\xi}^N), \Gamma_1(\bs{\xi}^N), \ldots, \Gamma_{n-1}(\bs{\xi}^N))$ is generally called the estimator and yields an estimate of
the unknown parameters, i.e., it is expected that $\bs{\Gamma}(\bs{\xi}^N) \approx \bs{\gamma}$.  Since the outcomes of the measurements fluctuate from run to run, the
estimation of the unknown parameters is always accompanied by an uncertainty.  As an example of an estimator, the maximum likelihood estimator is
defined as the value of $\bs{\Gamma}$ that maximizes $P(\bs{\xi}^N|\bs{\Gamma})$.

%

The performance of a multiparameter estimation is conveniently represented by a covariance matrix $\mathbf{C}$, whose matrix elements are given by
\begin{eqnarray}
  C_{i,j} &\equiv&  \av{ (\Gamma_i(\bs{\xi}^N) - \gamma_i\big)\big(\Gamma_j(\bs{\xi}^N) - \gamma_j\big)  }_P \nonumber \\
  &=& \sum_{\bs{\xi}^N} \big(\Gamma_i(\bs{\xi}^N) - \gamma_i\big)\big(\Gamma_j(\bs{\xi}^N) - \gamma_j\big) P(\bs{\xi}^N|\bs{\gamma}).
\label{Cmat}
\end{eqnarray}
Here the average in the first line is taken over the joint distribution $P(\bs{\xi}^N)$ and the sum over $\bs{\xi}^N$ in the second line is taken over all possible values of the
outcomes of $\bs{\xi}^N$.  For an unbiased estimator, the mean value of $\bs{\Gamma}(\bs{\xi}^N)$, taken over the probability distribution $P(\bs{\xi}^N)$, 
is equal to the true value of the unknown parameter vector $\bs{\gamma}$. In this case, the covariance matrix $\mathbf{C}$ satisfies the  inequality
\begin{equation}
\label{crlbCF}
  \mathbf{C} \geqslant \frac{1}{N} \mathbf{F}^{-1},
\end{equation}
which is known as the Cram\'er-Rao theorem~\cite{helstrom1976quantum, refregier2012noise} 
and it has to be understood in the matrix sense, i.e. $\mathbf{C} - N^{-1} \mathbf{F}^{-1}$ is a positive semi-definite matrix. The inequality~(\ref{crlbCF}) provides a (lower) bound for the performance of the multiparameter estimation and it is expressed in terms of the
inverse of the Fisher information matrix (FIM)~$\mathbf{F}$, whose elements are given by:
\begin{eqnarray}
\label{FIM}
  F_{i,j} = \sum_{\boldsymbol{\xi}} \frac{1}{p(\boldsymbol{\xi}| \boldsymbol{\gamma})} 
  \frac{ \partial p(\boldsymbol{\xi}| \boldsymbol{\gamma}) }{ \partial \gamma_i}
  \frac{ \partial p(\boldsymbol{\xi}| \boldsymbol{\gamma}) }{ \partial \gamma_j}.
\end{eqnarray}
The FIM depends only on the probability distribution $p(\bs{\xi}|\bs{\gamma})$ from which $P(\bs{\xi}^N)$ is constructed. The information about $P$ is inherited in the
statistical prefactor $1/N$, which ensures that smaller uncertainties can be attained when larger data sets are used for estimation.

In this work, we quantify the uncertainty $\Delta \gamma_i$ of the unknown parameter $\gamma_i$ by the variance of the estimator for each of the corresponding estimated parameters,
i.e., $(\Delta\gamma_i)^2 = [\mathbf{C}]_{i,i}$. The Cram\'er-Rao theorem states that such variances are bounded by the inverse of the Fisher information matrix~\cite{refregier2012noise}.
The bounds, which represent the minimal uncertainty 
that can be attained in the estimation of the parameters, take the form
\begin{equation}
\label{crlb}
  \Delta \gamma_i \geqslant \frac{1}{\sqrt N} [\mathbf{F}^{-1}]_{i,i}^{1/2}.
\end{equation}
These bounds are saturated asymptotically in $N$, in the limit of very large samples, by the maximum likelihood estimator~\cite{helstrom1976quantum, refregier2012noise}. 
We note, however, that in practice the estimated uncertainties will be larger than the above outlined bounds.

\bibliography{allarticles}

\begin{thebibliography}{61}%
\makeatletter
\providecommand \@ifxundefined [1]{%
 \@ifx{#1\undefined}
}%
\providecommand \@ifnum [1]{%
 \ifnum #1\expandafter \@firstoftwo
 \else \expandafter \@secondoftwo
 \fi
}%
\providecommand \@ifx [1]{%
 \ifx #1\expandafter \@firstoftwo
 \else \expandafter \@secondoftwo
 \fi
}%
\providecommand \natexlab [1]{#1}%
\providecommand \enquote  [1]{``#1''}%
\providecommand \bibnamefont  [1]{#1}%
\providecommand \bibfnamefont [1]{#1}%
\providecommand \citenamefont [1]{#1}%
\providecommand \href@noop [0]{\@secondoftwo}%
\providecommand \href [0]{\begingroup \@sanitize@url \@href}%
\providecommand \@href[1]{\@@startlink{#1}\@@href}%
\providecommand \@@href[1]{\endgroup#1\@@endlink}%
\providecommand \@sanitize@url [0]{\catcode `\\12\catcode `\$12\catcode
  `\&12\catcode `\#12\catcode `\^12\catcode `\_12\catcode `\%12\relax}%
\providecommand \@@startlink[1]{}%
\providecommand \@@endlink[0]{}%
\providecommand \url  [0]{\begingroup\@sanitize@url \@url }%
\providecommand \@url [1]{\endgroup\@href {#1}{\urlprefix }}%
\providecommand \urlprefix  [0]{URL }%
\providecommand \Eprint [0]{\href }%
\providecommand \doibase [0]{http://dx.doi.org/}%
\providecommand \selectlanguage [0]{\@gobble}%
\providecommand \bibinfo  [0]{\@secondoftwo}%
\providecommand \bibfield  [0]{\@secondoftwo}%
\providecommand \translation [1]{[#1]}%
\providecommand \BibitemOpen [0]{}%
\providecommand \bibitemStop [0]{}%
\providecommand \bibitemNoStop [0]{.\EOS\space}%
\providecommand \EOS [0]{\spacefactor3000\relax}%
\providecommand \BibitemShut  [1]{\csname bibitem#1\endcsname}%
\let\auto@bib@innerbib\@empty
\bibitem [{\citenamefont {Bloom}\ \emph {et~al.}(2014)\citenamefont {Bloom},
  \citenamefont {Nicholson}, \citenamefont {Williams} \emph
  {et~al.}}]{Bloom2014}%
  \BibitemOpen
  \bibfield  {author} {\bibinfo {author} {\bibfnamefont {B.}~\bibnamefont
  {Bloom}}, \bibinfo {author} {\bibfnamefont {T.}~\bibnamefont {Nicholson}},
  \bibinfo {author} {\bibfnamefont {J.}~\bibnamefont {Williams}},  \emph
  {et~al.},\ }\href@noop {} {\bibfield  {journal} {\bibinfo  {journal}
  {Nature}\ }\textbf {\bibinfo {volume} {506}},\ \bibinfo {pages} {71}
  (\bibinfo {year} {2014})}\BibitemShut {NoStop}%
\bibitem [{\citenamefont {Nicholson}\ \emph {et~al.}(2015)\citenamefont
  {Nicholson}, \citenamefont {Campbell}, \citenamefont {Hutson}, \citenamefont
  {Marti}, \citenamefont {Bloom}, \citenamefont {McNally}, \citenamefont
  {Zhang}, \citenamefont {Barrett}, \citenamefont {Safronova}, \citenamefont
  {Strouse} \emph {et~al.}}]{Nicholson2015}%
  \BibitemOpen
  \bibfield  {author} {\bibinfo {author} {\bibfnamefont {T.}~\bibnamefont
  {Nicholson}}, \bibinfo {author} {\bibfnamefont {S.}~\bibnamefont {Campbell}},
  \bibinfo {author} {\bibfnamefont {R.}~\bibnamefont {Hutson}}, \bibinfo
  {author} {\bibfnamefont {G.}~\bibnamefont {Marti}}, \bibinfo {author}
  {\bibfnamefont {B.}~\bibnamefont {Bloom}}, \bibinfo {author} {\bibfnamefont
  {R.}~\bibnamefont {McNally}}, \bibinfo {author} {\bibfnamefont
  {W.}~\bibnamefont {Zhang}}, \bibinfo {author} {\bibfnamefont
  {M.}~\bibnamefont {Barrett}}, \bibinfo {author} {\bibfnamefont
  {M.}~\bibnamefont {Safronova}}, \bibinfo {author} {\bibfnamefont
  {G.}~\bibnamefont {Strouse}},  \emph {et~al.},\ }\href@noop {} {\bibfield
  {journal} {\bibinfo  {journal} {Nature Communications}\ }\textbf {\bibinfo
  {volume} {6}} (\bibinfo {year} {2015})}\BibitemShut {NoStop}%
\bibitem [{\citenamefont {Hoover}\ \emph {et~al.}(1996)\citenamefont {Hoover},
  \citenamefont {Klein},\ and\ \citenamefont {Campbell}}]{Hoover1996}%
  \BibitemOpen
  \bibfield  {author} {\bibinfo {author} {\bibfnamefont {D.~B.}\ \bibnamefont
  {Hoover}}, \bibinfo {author} {\bibfnamefont {D.~P.}\ \bibnamefont {Klein}}, \
  and\ \bibinfo {author} {\bibfnamefont {D.~C.}\ \bibnamefont {Campbell}},\
  }\href@noop {} {\bibfield  {journal} {\bibinfo  {journal} {Report No.
  95-831}\ ,\ \bibinfo {pages} {U.S. Geological Survey}} (\bibinfo {year}
  {1996})}\BibitemShut {NoStop}%
\bibitem [{\citenamefont {DeGregoria}(2010)}]{DeGregoria2010}%
  \BibitemOpen
  \bibfield  {author} {\bibinfo {author} {\bibfnamefont {A.}~\bibnamefont
  {DeGregoria}},\ }\href@noop {} {\emph {\bibinfo {title} {Ph.D. thesis}}}\
  (\bibinfo  {publisher} {Air Force Institute of Technology},\ \bibinfo {year}
  {2010})\BibitemShut {NoStop}%
\bibitem [{\citenamefont {Sander}\ \emph {et~al.}(2012)\citenamefont {Sander},
  \citenamefont {Preusser}, \citenamefont {Mhaskar}, \citenamefont {Kitching},
  \citenamefont {Trahms},\ and\ \citenamefont {Knappe}}]{Sander2012}%
  \BibitemOpen
  \bibfield  {author} {\bibinfo {author} {\bibfnamefont {T.}~\bibnamefont
  {Sander}}, \bibinfo {author} {\bibfnamefont {J.}~\bibnamefont {Preusser}},
  \bibinfo {author} {\bibfnamefont {R.}~\bibnamefont {Mhaskar}}, \bibinfo
  {author} {\bibfnamefont {J.}~\bibnamefont {Kitching}}, \bibinfo {author}
  {\bibfnamefont {L.}~\bibnamefont {Trahms}}, \ and\ \bibinfo {author}
  {\bibfnamefont {S.}~\bibnamefont {Knappe}},\ }\href
  {https://www.osapublishing.org/boe/abstract.cfm?uri=boe-3-5-981} {\bibfield
  {journal} {\bibinfo  {journal} {Biomed. Opt. Express}\ }\textbf {\bibinfo
  {volume} {3}},\ \bibinfo {pages} {981} (\bibinfo {year} {2012})}\BibitemShut
  {NoStop}%
\bibitem [{\citenamefont {Jensen}\ \emph {et~al.}(2016)\citenamefont {Jensen},
  \citenamefont {Budvytyte}, \citenamefont {Thomas}, \citenamefont {Wang},
  \citenamefont {Fuchs}, \citenamefont {Balabas}, \citenamefont {Vasilakis},
  \citenamefont {Mosgaard}, \citenamefont {St{\ae}rkind}, \citenamefont
  {M\"uller}, \citenamefont {Heimburg}, \citenamefont {Olesen},\ and\
  \citenamefont {Polzik}}]{Jensen2016}%
  \BibitemOpen
  \bibfield  {author} {\bibinfo {author} {\bibfnamefont {K.}~\bibnamefont
  {Jensen}}, \bibinfo {author} {\bibfnamefont {R.}~\bibnamefont {Budvytyte}},
  \bibinfo {author} {\bibfnamefont {R.~A.}\ \bibnamefont {Thomas}}, \bibinfo
  {author} {\bibfnamefont {T.}~\bibnamefont {Wang}}, \bibinfo {author}
  {\bibfnamefont {A.~M.}\ \bibnamefont {Fuchs}}, \bibinfo {author}
  {\bibfnamefont {M.~V.}\ \bibnamefont {Balabas}}, \bibinfo {author}
  {\bibfnamefont {G.}~\bibnamefont {Vasilakis}}, \bibinfo {author}
  {\bibfnamefont {L.~D.}\ \bibnamefont {Mosgaard}}, \bibinfo {author}
  {\bibfnamefont {H.~C.}\ \bibnamefont {St{\ae}rkind}}, \bibinfo {author}
  {\bibfnamefont {J.~H.}\ \bibnamefont {M\"uller}}, \bibinfo {author}
  {\bibfnamefont {T.}~\bibnamefont {Heimburg}}, \bibinfo {author}
  {\bibfnamefont {S.-P.}\ \bibnamefont {Olesen}}, \ and\ \bibinfo {author}
  {\bibfnamefont {E.~S.}\ \bibnamefont {Polzik}},\ }\href
  {https://www.nature.com/articles/srep29638} {\bibfield  {journal} {\bibinfo
  {journal} {Sci. Rep.}\ }\textbf {\bibinfo {volume} {6}},\ \bibinfo {pages}
  {29638} (\bibinfo {year} {2016})}\BibitemShut {NoStop}%
\bibitem [{\citenamefont {Balasubramanian}\ \emph {et~al.}(2008)\citenamefont
  {Balasubramanian}, \citenamefont {Chan}, \citenamefont {Kolesov},
  \citenamefont {Al-Hmoud}, \citenamefont {Tisler}, \citenamefont {Shin},
  \citenamefont {Kim}, \citenamefont {Wojcik}, \citenamefont {Hemmer},
  \citenamefont {Krueger} \emph {et~al.}}]{Balasubramanian2008}%
  \BibitemOpen
  \bibfield  {author} {\bibinfo {author} {\bibfnamefont {G.}~\bibnamefont
  {Balasubramanian}}, \bibinfo {author} {\bibfnamefont {I.}~\bibnamefont
  {Chan}}, \bibinfo {author} {\bibfnamefont {R.}~\bibnamefont {Kolesov}},
  \bibinfo {author} {\bibfnamefont {M.}~\bibnamefont {Al-Hmoud}}, \bibinfo
  {author} {\bibfnamefont {J.}~\bibnamefont {Tisler}}, \bibinfo {author}
  {\bibfnamefont {C.}~\bibnamefont {Shin}}, \bibinfo {author} {\bibfnamefont
  {C.}~\bibnamefont {Kim}}, \bibinfo {author} {\bibfnamefont {A.}~\bibnamefont
  {Wojcik}}, \bibinfo {author} {\bibfnamefont {P.~R.}\ \bibnamefont {Hemmer}},
  \bibinfo {author} {\bibfnamefont {A.}~\bibnamefont {Krueger}},  \emph
  {et~al.},\ }\href@noop {} {\bibfield  {journal} {\bibinfo  {journal}
  {Nature}\ }\textbf {\bibinfo {volume} {455}},\ \bibinfo {pages} {648}
  (\bibinfo {year} {2008})}\BibitemShut {NoStop}%
\bibitem [{\citenamefont {Chen}\ \emph {et~al.}(2006)\citenamefont {Chen},
  \citenamefont {Wilson},\ and\ \citenamefont {Tapley}}]{Chen2006}%
  \BibitemOpen
  \bibfield  {author} {\bibinfo {author} {\bibfnamefont {J.~L.}\ \bibnamefont
  {Chen}}, \bibinfo {author} {\bibfnamefont {C.~R.}\ \bibnamefont {Wilson}}, \
  and\ \bibinfo {author} {\bibfnamefont {B.~D.}\ \bibnamefont {Tapley}},\
  }\href@noop {} {\bibfield  {journal} {\bibinfo  {journal} {Science}\ }\textbf
  {\bibinfo {volume} {313}},\ \bibinfo {pages} {1958} (\bibinfo {year}
  {2006})}\BibitemShut {NoStop}%
\bibitem [{\citenamefont {Webb}\ \emph {et~al.}(1999)\citenamefont {Webb},
  \citenamefont {Flambaum}, \citenamefont {Churchill}, \citenamefont
  {Drinkwater},\ and\ \citenamefont {Barrow}}]{Webb1999}%
  \BibitemOpen
  \bibfield  {author} {\bibinfo {author} {\bibfnamefont {J.~K.}\ \bibnamefont
  {Webb}}, \bibinfo {author} {\bibfnamefont {V.~V.}\ \bibnamefont {Flambaum}},
  \bibinfo {author} {\bibfnamefont {C.~W.}\ \bibnamefont {Churchill}}, \bibinfo
  {author} {\bibfnamefont {M.~J.}\ \bibnamefont {Drinkwater}}, \ and\ \bibinfo
  {author} {\bibfnamefont {J.~D.}\ \bibnamefont {Barrow}},\ }\href@noop {}
  {\bibfield  {journal} {\bibinfo  {journal} {Phys. Rev. Lett.}\ }\textbf
  {\bibinfo {volume} {82}},\ \bibinfo {pages} {884} (\bibinfo {year}
  {1999})}\BibitemShut {NoStop}%
\bibitem [{\citenamefont {Chin}\ and\ \citenamefont
  {Flambaum}(2006)}]{Chin2006}%
  \BibitemOpen
  \bibfield  {author} {\bibinfo {author} {\bibfnamefont {C.}~\bibnamefont
  {Chin}}\ and\ \bibinfo {author} {\bibfnamefont {V.~V.}\ \bibnamefont
  {Flambaum}},\ }\href@noop {} {\bibfield  {journal} {\bibinfo  {journal}
  {Phys. Rev. Lett.}\ }\textbf {\bibinfo {volume} {96}},\ \bibinfo {pages}
  {230801} (\bibinfo {year} {2006})}\BibitemShut {NoStop}%
\bibitem [{\citenamefont {Zelevinsky}\ \emph {et~al.}(2008)\citenamefont
  {Zelevinsky}, \citenamefont {Kotochigova},\ and\ \citenamefont
  {Ye}}]{Zelevinsky2008}%
  \BibitemOpen
  \bibfield  {author} {\bibinfo {author} {\bibfnamefont {T.}~\bibnamefont
  {Zelevinsky}}, \bibinfo {author} {\bibfnamefont {S.}~\bibnamefont
  {Kotochigova}}, \ and\ \bibinfo {author} {\bibfnamefont {J.}~\bibnamefont
  {Ye}},\ }\href@noop {} {\bibfield  {journal} {\bibinfo  {journal} {Phys. Rev.
  Lett.}\ }\textbf {\bibinfo {volume} {100}},\ \bibinfo {pages} {043201}
  (\bibinfo {year} {2008})}\BibitemShut {NoStop}%
\bibitem [{\citenamefont {Blatt}\ \emph {et~al.}(2008)\citenamefont {Blatt},
  \citenamefont {Ludlow}, \citenamefont {Campbell}, \citenamefont {Thomsen},
  \citenamefont {Zelevinsky}, \citenamefont {Boyd}, \citenamefont {Ye},
  \citenamefont {Baillard}, \citenamefont {Fouch{\'e}}, \citenamefont
  {Le~Targat} \emph {et~al.}}]{Blatt2008}%
  \BibitemOpen
  \bibfield  {author} {\bibinfo {author} {\bibfnamefont {S.}~\bibnamefont
  {Blatt}}, \bibinfo {author} {\bibfnamefont {A.}~\bibnamefont {Ludlow}},
  \bibinfo {author} {\bibfnamefont {G.}~\bibnamefont {Campbell}}, \bibinfo
  {author} {\bibfnamefont {J.~W.}\ \bibnamefont {Thomsen}}, \bibinfo {author}
  {\bibfnamefont {T.}~\bibnamefont {Zelevinsky}}, \bibinfo {author}
  {\bibfnamefont {M.}~\bibnamefont {Boyd}}, \bibinfo {author} {\bibfnamefont
  {J.}~\bibnamefont {Ye}}, \bibinfo {author} {\bibfnamefont {X.}~\bibnamefont
  {Baillard}}, \bibinfo {author} {\bibfnamefont {M.}~\bibnamefont
  {Fouch{\'e}}}, \bibinfo {author} {\bibfnamefont {R.}~\bibnamefont
  {Le~Targat}},  \emph {et~al.},\ }\href@noop {} {\bibfield  {journal}
  {\bibinfo  {journal} {Phys. Rev. Lett.}\ }\textbf {\bibinfo {volume} {100}},\
  \bibinfo {pages} {140801} (\bibinfo {year} {2008})}\BibitemShut {NoStop}%
\bibitem [{\citenamefont {Hudson}\ \emph {et~al.}(2011)\citenamefont {Hudson},
  \citenamefont {Kara}, \citenamefont {Smallman}, \citenamefont {Sauer},
  \citenamefont {Tarbutt},\ and\ \citenamefont {Hinds}}]{Hudson2011}%
  \BibitemOpen
  \bibfield  {author} {\bibinfo {author} {\bibfnamefont {J.~J.}\ \bibnamefont
  {Hudson}}, \bibinfo {author} {\bibfnamefont {D.~M.}\ \bibnamefont {Kara}},
  \bibinfo {author} {\bibfnamefont {I.}~\bibnamefont {Smallman}}, \bibinfo
  {author} {\bibfnamefont {B.~E.}\ \bibnamefont {Sauer}}, \bibinfo {author}
  {\bibfnamefont {M.~R.}\ \bibnamefont {Tarbutt}}, \ and\ \bibinfo {author}
  {\bibfnamefont {E.~A.}\ \bibnamefont {Hinds}},\ }\href@noop {} {\bibfield
  {journal} {\bibinfo  {journal} {Nature}\ }\textbf {\bibinfo {volume} {473}},\
  \bibinfo {pages} {493} (\bibinfo {year} {2011})}\BibitemShut {NoStop}%
\bibitem [{\citenamefont {Baron}\ \emph {et~al.}(2014)\citenamefont {Baron},
  \citenamefont {Campbell}, \citenamefont {DeMille}, \citenamefont {Doyle},
  \citenamefont {Gabrielse}, \citenamefont {Gurevich}, \citenamefont {Hess},
  \citenamefont {Hutzler}, \citenamefont {Kirilov}, \citenamefont {Kozyryev}
  \emph {et~al.}}]{Baron2014}%
  \BibitemOpen
  \bibfield  {author} {\bibinfo {author} {\bibfnamefont {J.}~\bibnamefont
  {Baron}}, \bibinfo {author} {\bibfnamefont {W.~C.}\ \bibnamefont {Campbell}},
  \bibinfo {author} {\bibfnamefont {D.}~\bibnamefont {DeMille}}, \bibinfo
  {author} {\bibfnamefont {J.~M.}\ \bibnamefont {Doyle}}, \bibinfo {author}
  {\bibfnamefont {G.}~\bibnamefont {Gabrielse}}, \bibinfo {author}
  {\bibfnamefont {Y.~V.}\ \bibnamefont {Gurevich}}, \bibinfo {author}
  {\bibfnamefont {P.~W.}\ \bibnamefont {Hess}}, \bibinfo {author}
  {\bibfnamefont {N.~R.}\ \bibnamefont {Hutzler}}, \bibinfo {author}
  {\bibfnamefont {E.}~\bibnamefont {Kirilov}}, \bibinfo {author} {\bibfnamefont
  {I.}~\bibnamefont {Kozyryev}},  \emph {et~al.},\ }\href@noop {} {\bibfield
  {journal} {\bibinfo  {journal} {Science}\ }\textbf {\bibinfo {volume}
  {343}},\ \bibinfo {pages} {269} (\bibinfo {year} {2014})}\BibitemShut
  {NoStop}%
\bibitem [{\citenamefont {Schnabel}\ \emph {et~al.}(2010)\citenamefont
  {Schnabel}, \citenamefont {Mavalvala}, \citenamefont {McClelland},\ and\
  \citenamefont {Lam}}]{Schnabel2010}%
  \BibitemOpen
  \bibfield  {author} {\bibinfo {author} {\bibfnamefont {R.}~\bibnamefont
  {Schnabel}}, \bibinfo {author} {\bibfnamefont {N.}~\bibnamefont {Mavalvala}},
  \bibinfo {author} {\bibfnamefont {D.~E.}\ \bibnamefont {McClelland}}, \ and\
  \bibinfo {author} {\bibfnamefont {P.~K.}\ \bibnamefont {Lam}},\ }\href@noop
  {} {\bibfield  {journal} {\bibinfo  {journal} {Nat. Comm.}\ }\textbf
  {\bibinfo {volume} {1}},\ \bibinfo {pages} {121} (\bibinfo {year}
  {2010})}\BibitemShut {NoStop}%
\bibitem [{\citenamefont {Schlippert}\ \emph {et~al.}(2014)\citenamefont
  {Schlippert}, \citenamefont {Hartwig}, \citenamefont {Albers}, \citenamefont
  {Richardson}, \citenamefont {Schubert}, \citenamefont {Roura}, \citenamefont
  {Schleich}, \citenamefont {Ertmer},\ and\ \citenamefont
  {Rasel}}]{Schlippert2014}%
  \BibitemOpen
  \bibfield  {author} {\bibinfo {author} {\bibfnamefont {D.}~\bibnamefont
  {Schlippert}}, \bibinfo {author} {\bibfnamefont {J.}~\bibnamefont {Hartwig}},
  \bibinfo {author} {\bibfnamefont {H.}~\bibnamefont {Albers}}, \bibinfo
  {author} {\bibfnamefont {L.~L.}\ \bibnamefont {Richardson}}, \bibinfo
  {author} {\bibfnamefont {C.}~\bibnamefont {Schubert}}, \bibinfo {author}
  {\bibfnamefont {A.}~\bibnamefont {Roura}}, \bibinfo {author} {\bibfnamefont
  {W.~P.}\ \bibnamefont {Schleich}}, \bibinfo {author} {\bibfnamefont
  {W.}~\bibnamefont {Ertmer}}, \ and\ \bibinfo {author} {\bibfnamefont {E.~M.}\
  \bibnamefont {Rasel}},\ }\href
  {https://link.aps.org/doi/10.1103/PhysRevLett.112.203002} {\bibfield
  {journal} {\bibinfo  {journal} {Phys. Rev. Lett.}\ }\textbf {\bibinfo
  {volume} {112}},\ \bibinfo {pages} {203002} (\bibinfo {year}
  {2014})}\BibitemShut {NoStop}%
\bibitem [{\citenamefont {Biedermann}\ \emph {et~al.}(2015)\citenamefont
  {Biedermann}, \citenamefont {Wu}, \citenamefont {Deslauriers}, \citenamefont
  {Roy}, \citenamefont {Mahadeswaraswamy},\ and\ \citenamefont
  {Kasevich}}]{Biedermann2015}%
  \BibitemOpen
  \bibfield  {author} {\bibinfo {author} {\bibfnamefont {G.~W.}\ \bibnamefont
  {Biedermann}}, \bibinfo {author} {\bibfnamefont {X.}~\bibnamefont {Wu}},
  \bibinfo {author} {\bibfnamefont {L.}~\bibnamefont {Deslauriers}}, \bibinfo
  {author} {\bibfnamefont {S.}~\bibnamefont {Roy}}, \bibinfo {author}
  {\bibfnamefont {C.}~\bibnamefont {Mahadeswaraswamy}}, \ and\ \bibinfo
  {author} {\bibfnamefont {M.~A.}\ \bibnamefont {Kasevich}},\ }\href {\doibase
  10.1103/PhysRevA.91.033629} {\bibfield  {journal} {\bibinfo  {journal} {Phys.
  Rev. A}\ }\textbf {\bibinfo {volume} {91}},\ \bibinfo {pages} {033629}
  (\bibinfo {year} {2015})}\BibitemShut {NoStop}%
\bibitem [{\citenamefont {Rosi}\ \emph {et~al.}(2017)\citenamefont {Rosi},
  \citenamefont {D'Amico}, \citenamefont {Cacciapuoti}, \citenamefont
  {Sorrentino}, \citenamefont {Prevedelli}, \citenamefont {Zych}, \citenamefont
  {Brukner},\ and\ \citenamefont {Tino}}]{Rosi2017}%
  \BibitemOpen
  \bibfield  {author} {\bibinfo {author} {\bibfnamefont {G.}~\bibnamefont
  {Rosi}}, \bibinfo {author} {\bibfnamefont {G.}~\bibnamefont {D'Amico}},
  \bibinfo {author} {\bibfnamefont {L.}~\bibnamefont {Cacciapuoti}}, \bibinfo
  {author} {\bibfnamefont {F.}~\bibnamefont {Sorrentino}}, \bibinfo {author}
  {\bibfnamefont {M.}~\bibnamefont {Prevedelli}}, \bibinfo {author}
  {\bibfnamefont {M.}~\bibnamefont {Zych}}, \bibinfo {author} {\bibfnamefont
  {{\v{C}}.}~\bibnamefont {Brukner}}, \ and\ \bibinfo {author} {\bibfnamefont
  {G.~M.}\ \bibnamefont {Tino}},\ }\href@noop {} {\bibfield  {journal}
  {\bibinfo  {journal} {Nat. Comm.}\ }\textbf {\bibinfo {volume} {8}},\
  \bibinfo {pages} {15529} (\bibinfo {year} {2017})}\BibitemShut {NoStop}%
\bibitem [{\citenamefont {Ferrari}\ \emph {et~al.}(2006)\citenamefont
  {Ferrari}, \citenamefont {Poli}, \citenamefont {Sorrentino},\ and\
  \citenamefont {Tino}}]{Ferrari2006}%
  \BibitemOpen
  \bibfield  {author} {\bibinfo {author} {\bibfnamefont {G.}~\bibnamefont
  {Ferrari}}, \bibinfo {author} {\bibfnamefont {N.}~\bibnamefont {Poli}},
  \bibinfo {author} {\bibfnamefont {F.}~\bibnamefont {Sorrentino}}, \ and\
  \bibinfo {author} {\bibfnamefont {G.~M.}\ \bibnamefont {Tino}},\ }\href
  {https://link.aps.org/doi/10.1103/PhysRevLett.97.060402} {\bibfield
  {journal} {\bibinfo  {journal} {Phys. Rev. Lett.}\ }\textbf {\bibinfo
  {volume} {97}},\ \bibinfo {pages} {060402} (\bibinfo {year}
  {2006})}\BibitemShut {NoStop}%
\bibitem [{\citenamefont {Salumbides}\ \emph {et~al.}(2013)\citenamefont
  {Salumbides}, \citenamefont {Koelemeij}, \citenamefont {Komasa},
  \citenamefont {Pachucki}, \citenamefont {Eikema},\ and\ \citenamefont
  {Ubachs}}]{Salumbides2013}%
  \BibitemOpen
  \bibfield  {author} {\bibinfo {author} {\bibfnamefont {E.~J.}\ \bibnamefont
  {Salumbides}}, \bibinfo {author} {\bibfnamefont {J.~C.~J.}\ \bibnamefont
  {Koelemeij}}, \bibinfo {author} {\bibfnamefont {J.}~\bibnamefont {Komasa}},
  \bibinfo {author} {\bibfnamefont {K.}~\bibnamefont {Pachucki}}, \bibinfo
  {author} {\bibfnamefont {K.~S.~E.}\ \bibnamefont {Eikema}}, \ and\ \bibinfo
  {author} {\bibfnamefont {W.}~\bibnamefont {Ubachs}},\ }\href {\doibase
  10.1103/PhysRevD.87.112008} {\bibfield  {journal} {\bibinfo  {journal} {Phys.
  Rev. D}\ }\textbf {\bibinfo {volume} {87}},\ \bibinfo {pages} {112008}
  (\bibinfo {year} {2013})}\BibitemShut {NoStop}%
\bibitem [{\citenamefont {Borkowski}\ \emph {et~al.}(2017)\citenamefont
  {Borkowski}, \citenamefont {Buchachenko}, \citenamefont {Ciury{\l}o},
  \citenamefont {Julienne}, \citenamefont {Yamada}, \citenamefont {Yuu},
  \citenamefont {Takahashi}, \citenamefont {Takasu},\ and\ \citenamefont
  {Takahashi}}]{Borkowski2017}%
  \BibitemOpen
  \bibfield  {author} {\bibinfo {author} {\bibfnamefont {M.}~\bibnamefont
  {Borkowski}}, \bibinfo {author} {\bibfnamefont {A.}~\bibnamefont
  {Buchachenko}}, \bibinfo {author} {\bibfnamefont {R.}~\bibnamefont
  {Ciury{\l}o}}, \bibinfo {author} {\bibfnamefont {P.}~\bibnamefont
  {Julienne}}, \bibinfo {author} {\bibfnamefont {H.}~\bibnamefont {Yamada}},
  \bibinfo {author} {\bibfnamefont {K.}~\bibnamefont {Yuu}}, \bibinfo {author}
  {\bibfnamefont {K.}~\bibnamefont {Takahashi}}, \bibinfo {author}
  {\bibfnamefont {Y.}~\bibnamefont {Takasu}}, \ and\ \bibinfo {author}
  {\bibfnamefont {Y.}~\bibnamefont {Takahashi}},\ }in\ \href@noop {} {\emph
  {\bibinfo {booktitle} {Journal of Physics: Conference Series}}},\ Vol.\
  \bibinfo {volume} {810}\ (\bibinfo {organization} {IOP Publishing},\ \bibinfo
  {year} {2017})\ p.\ \bibinfo {pages} {012014}\BibitemShut {NoStop}%
\bibitem [{\citenamefont {Giovannetti}\ \emph {et~al.}(2004)\citenamefont
  {Giovannetti}, \citenamefont {Lloyd},\ and\ \citenamefont
  {Maccone}}]{Giovannetti2004}%
  \BibitemOpen
  \bibfield  {author} {\bibinfo {author} {\bibfnamefont {V.}~\bibnamefont
  {Giovannetti}}, \bibinfo {author} {\bibfnamefont {S.}~\bibnamefont {Lloyd}},
  \ and\ \bibinfo {author} {\bibfnamefont {L.}~\bibnamefont {Maccone}},\
  }\href@noop {} {\bibfield  {journal} {\bibinfo  {journal} {Science}\ }\textbf
  {\bibinfo {volume} {306}},\ \bibinfo {pages} {1330} (\bibinfo {year}
  {2004})}\BibitemShut {NoStop}%
\bibitem [{\citenamefont {Giovannetti}\ \emph {et~al.}(2006)\citenamefont
  {Giovannetti}, \citenamefont {Lloyd},\ and\ \citenamefont
  {Maccone}}]{Giovannetti2006}%
  \BibitemOpen
  \bibfield  {author} {\bibinfo {author} {\bibfnamefont {V.}~\bibnamefont
  {Giovannetti}}, \bibinfo {author} {\bibfnamefont {S.}~\bibnamefont {Lloyd}},
  \ and\ \bibinfo {author} {\bibfnamefont {L.}~\bibnamefont {Maccone}},\ }\href
  {\doibase 10.1103/PhysRevLett.96.010401} {\bibfield  {journal} {\bibinfo
  {journal} {Phys. Rev. Lett.}\ }\textbf {\bibinfo {volume} {96}},\ \bibinfo
  {pages} {010401} (\bibinfo {year} {2006})}\BibitemShut {NoStop}%
\bibitem [{\citenamefont {Petersen}\ \emph {et~al.}(2005)\citenamefont
  {Petersen}, \citenamefont {Madsen},\ and\ \citenamefont
  {M\o{}lmer}}]{Petersen2005}%
  \BibitemOpen
  \bibfield  {author} {\bibinfo {author} {\bibfnamefont {V.}~\bibnamefont
  {Petersen}}, \bibinfo {author} {\bibfnamefont {L.~B.}\ \bibnamefont
  {Madsen}}, \ and\ \bibinfo {author} {\bibfnamefont {K.}~\bibnamefont
  {M\o{}lmer}},\ }\href {https://link.aps.org/doi/10.1103/PhysRevA.71.012312}
  {\bibfield  {journal} {\bibinfo  {journal} {Phys. Rev. A}\ }\textbf {\bibinfo
  {volume} {71}},\ \bibinfo {pages} {012312} (\bibinfo {year}
  {2005})}\BibitemShut {NoStop}%
\bibitem [{\citenamefont {Petersen}\ and\ \citenamefont
  {M\o{}lmer}(2006)}]{Petersen2006}%
  \BibitemOpen
  \bibfield  {author} {\bibinfo {author} {\bibfnamefont {V.}~\bibnamefont
  {Petersen}}\ and\ \bibinfo {author} {\bibfnamefont {K.}~\bibnamefont
  {M\o{}lmer}},\ }\href {https://link.aps.org/doi/10.1103/PhysRevA.74.043802}
  {\bibfield  {journal} {\bibinfo  {journal} {Phys. Rev. A}\ }\textbf {\bibinfo
  {volume} {74}},\ \bibinfo {pages} {043802} (\bibinfo {year}
  {2006})}\BibitemShut {NoStop}%
\bibitem [{\citenamefont {Chase}\ and\ \citenamefont
  {Geremia}(2009)}]{Chase2009a}%
  \BibitemOpen
  \bibfield  {author} {\bibinfo {author} {\bibfnamefont {B.~A.}\ \bibnamefont
  {Chase}}\ and\ \bibinfo {author} {\bibfnamefont {J.~M.}\ \bibnamefont
  {Geremia}},\ }\href {https://link.aps.org/doi/10.1103/PhysRevA.79.022314}
  {\bibfield  {journal} {\bibinfo  {journal} {Phys. Rev. A}\ }\textbf {\bibinfo
  {volume} {79}},\ \bibinfo {pages} {022314} (\bibinfo {year}
  {2009})}\BibitemShut {NoStop}%
\bibitem [{\citenamefont {Chase}\ \emph {et~al.}(2009)\citenamefont {Chase},
  \citenamefont {Baragiola}, \citenamefont {Partner}, \citenamefont {Black},\
  and\ \citenamefont {Geremia}}]{Chase2009b}%
  \BibitemOpen
  \bibfield  {author} {\bibinfo {author} {\bibfnamefont {B.~A.}\ \bibnamefont
  {Chase}}, \bibinfo {author} {\bibfnamefont {B.~Q.}\ \bibnamefont
  {Baragiola}}, \bibinfo {author} {\bibfnamefont {H.~L.}\ \bibnamefont
  {Partner}}, \bibinfo {author} {\bibfnamefont {B.~D.}\ \bibnamefont {Black}},
  \ and\ \bibinfo {author} {\bibfnamefont {J.~M.}\ \bibnamefont {Geremia}},\
  }\href {https://link.aps.org/doi/10.1103/PhysRevA.79.062107} {\bibfield
  {journal} {\bibinfo  {journal} {Phys. Rev. A}\ }\textbf {\bibinfo {volume}
  {79}},\ \bibinfo {pages} {062107} (\bibinfo {year} {2009})}\BibitemShut
  {NoStop}%
\bibitem [{\citenamefont {Negretti}\ and\ \citenamefont
  {M{\o}lmer}(2013)}]{Negretti2013}%
  \BibitemOpen
  \bibfield  {author} {\bibinfo {author} {\bibfnamefont {A.}~\bibnamefont
  {Negretti}}\ and\ \bibinfo {author} {\bibfnamefont {K.}~\bibnamefont
  {M{\o}lmer}},\ }\href@noop {} {\bibfield  {journal} {\bibinfo  {journal} {New
  Journal of Physics}\ }\textbf {\bibinfo {volume} {15}},\ \bibinfo {pages}
  {125002} (\bibinfo {year} {2013})}\BibitemShut {NoStop}%
\bibitem [{\citenamefont {Martin~Ciurana}\ \emph {et~al.}(2017)\citenamefont
  {Martin~Ciurana}, \citenamefont {Colangelo}, \citenamefont
  {Slodi\ifmmode~\check{c}\else \v{c}\fi{}ka}, \citenamefont {Sewell},\ and\
  \citenamefont {Mitchell}}]{Ciurana2017}%
  \BibitemOpen
  \bibfield  {author} {\bibinfo {author} {\bibfnamefont {F.}~\bibnamefont
  {Martin~Ciurana}}, \bibinfo {author} {\bibfnamefont {G.}~\bibnamefont
  {Colangelo}}, \bibinfo {author} {\bibfnamefont {L.}~\bibnamefont
  {Slodi\ifmmode~\check{c}\else \v{c}\fi{}ka}}, \bibinfo {author}
  {\bibfnamefont {R.~J.}\ \bibnamefont {Sewell}}, \ and\ \bibinfo {author}
  {\bibfnamefont {M.~W.}\ \bibnamefont {Mitchell}},\ }\href {\doibase
  10.1103/PhysRevLett.119.043603} {\bibfield  {journal} {\bibinfo  {journal}
  {Phys. Rev. Lett.}\ }\textbf {\bibinfo {volume} {119}},\ \bibinfo {pages}
  {043603} (\bibinfo {year} {2017})}\BibitemShut {NoStop}%
\bibitem [{\citenamefont {Fernholz}\ \emph {et~al.}(2008)\citenamefont
  {Fernholz}, \citenamefont {Krauter}, \citenamefont {Jensen}, \citenamefont
  {Sherson}, \citenamefont {S\o{}rensen},\ and\ \citenamefont
  {Polzik}}]{Fernholz2008}%
  \BibitemOpen
  \bibfield  {author} {\bibinfo {author} {\bibfnamefont {T.}~\bibnamefont
  {Fernholz}}, \bibinfo {author} {\bibfnamefont {H.}~\bibnamefont {Krauter}},
  \bibinfo {author} {\bibfnamefont {K.}~\bibnamefont {Jensen}}, \bibinfo
  {author} {\bibfnamefont {J.~F.}\ \bibnamefont {Sherson}}, \bibinfo {author}
  {\bibfnamefont {A.~S.}\ \bibnamefont {S\o{}rensen}}, \ and\ \bibinfo {author}
  {\bibfnamefont {E.~S.}\ \bibnamefont {Polzik}},\ }\href
  {https://link.aps.org/doi/10.1103/PhysRevLett.101.073601} {\bibfield
  {journal} {\bibinfo  {journal} {Phys. Rev. Lett.}\ }\textbf {\bibinfo
  {volume} {101}},\ \bibinfo {pages} {073601} (\bibinfo {year}
  {2008})}\BibitemShut {NoStop}%
\bibitem [{\citenamefont {Wasilewski}\ \emph {et~al.}(2010)\citenamefont
  {Wasilewski}, \citenamefont {Jensen}, \citenamefont {Krauter}, \citenamefont
  {Renema}, \citenamefont {Balabas},\ and\ \citenamefont
  {Polzik}}]{Wasilewski2010}%
  \BibitemOpen
  \bibfield  {author} {\bibinfo {author} {\bibfnamefont {W.}~\bibnamefont
  {Wasilewski}}, \bibinfo {author} {\bibfnamefont {K.}~\bibnamefont {Jensen}},
  \bibinfo {author} {\bibfnamefont {H.}~\bibnamefont {Krauter}}, \bibinfo
  {author} {\bibfnamefont {J.~J.}\ \bibnamefont {Renema}}, \bibinfo {author}
  {\bibfnamefont {M.~V.}\ \bibnamefont {Balabas}}, \ and\ \bibinfo {author}
  {\bibfnamefont {E.~S.}\ \bibnamefont {Polzik}},\ }\href
  {https://link.aps.org/doi/10.1103/PhysRevLett.104.133601} {\bibfield
  {journal} {\bibinfo  {journal} {Phys. Rev. Lett.}\ }\textbf {\bibinfo
  {volume} {104}},\ \bibinfo {pages} {133601} (\bibinfo {year}
  {2010})}\BibitemShut {NoStop}%
\bibitem [{\citenamefont {Riedel}\ \emph {et~al.}(2010)\citenamefont {Riedel},
  \citenamefont {B\"ohi}, \citenamefont {Li}, \citenamefont {H\"ansch},
  \citenamefont {Sinatra},\ and\ \citenamefont {Treutlein}}]{Riedel2010}%
  \BibitemOpen
  \bibfield  {author} {\bibinfo {author} {\bibfnamefont {M.~F.}\ \bibnamefont
  {Riedel}}, \bibinfo {author} {\bibfnamefont {P.}~\bibnamefont {B\"ohi}},
  \bibinfo {author} {\bibfnamefont {Y.}~\bibnamefont {Li}}, \bibinfo {author}
  {\bibfnamefont {T.~W.}\ \bibnamefont {H\"ansch}}, \bibinfo {author}
  {\bibfnamefont {A.}~\bibnamefont {Sinatra}}, \ and\ \bibinfo {author}
  {\bibfnamefont {P.}~\bibnamefont {Treutlein}},\ }\href@noop {} {\bibfield
  {journal} {\bibinfo  {journal} {Nature}\ }\textbf {\bibinfo {volume} {464}},\
  \bibinfo {pages} {1170} (\bibinfo {year} {2010})}\BibitemShut {NoStop}%
\bibitem [{\citenamefont {Krischek}\ \emph {et~al.}(2011)\citenamefont
  {Krischek}, \citenamefont {Schwemmer}, \citenamefont {Wieczorek},
  \citenamefont {Weinfurter}, \citenamefont {Hyllus}, \citenamefont {Pezz\'e},\
  and\ \citenamefont {Smerzi}}]{Krischek2011}%
  \BibitemOpen
  \bibfield  {author} {\bibinfo {author} {\bibfnamefont {R.}~\bibnamefont
  {Krischek}}, \bibinfo {author} {\bibfnamefont {C.}~\bibnamefont {Schwemmer}},
  \bibinfo {author} {\bibfnamefont {W.}~\bibnamefont {Wieczorek}}, \bibinfo
  {author} {\bibfnamefont {H.}~\bibnamefont {Weinfurter}}, \bibinfo {author}
  {\bibfnamefont {P.}~\bibnamefont {Hyllus}}, \bibinfo {author} {\bibfnamefont
  {L.}~\bibnamefont {Pezz\'e}}, \ and\ \bibinfo {author} {\bibfnamefont
  {A.}~\bibnamefont {Smerzi}},\ }\href
  {https://link.aps.org/doi/10.1103/PhysRevLett.107.080504} {\bibfield
  {journal} {\bibinfo  {journal} {Phys. Rev. Lett.}\ }\textbf {\bibinfo
  {volume} {107}},\ \bibinfo {pages} {080504} (\bibinfo {year}
  {2011})}\BibitemShut {NoStop}%
\bibitem [{\citenamefont {Hardman}\ \emph {et~al.}(2016)\citenamefont
  {Hardman}, \citenamefont {Everitt}, \citenamefont {McDonald}, \citenamefont
  {Manju}, \citenamefont {Wigley}, \citenamefont {Sooriyabandara},
  \citenamefont {Kuhn}, \citenamefont {Debs}, \citenamefont {Close},\ and\
  \citenamefont {Robins}}]{Hardman2016}%
  \BibitemOpen
  \bibfield  {author} {\bibinfo {author} {\bibfnamefont {K.~S.}\ \bibnamefont
  {Hardman}}, \bibinfo {author} {\bibfnamefont {P.~J.}\ \bibnamefont
  {Everitt}}, \bibinfo {author} {\bibfnamefont {G.~D.}\ \bibnamefont
  {McDonald}}, \bibinfo {author} {\bibfnamefont {P.}~\bibnamefont {Manju}},
  \bibinfo {author} {\bibfnamefont {P.~B.}\ \bibnamefont {Wigley}}, \bibinfo
  {author} {\bibfnamefont {M.~A.}\ \bibnamefont {Sooriyabandara}}, \bibinfo
  {author} {\bibfnamefont {C.~C.~N.}\ \bibnamefont {Kuhn}}, \bibinfo {author}
  {\bibfnamefont {J.~E.}\ \bibnamefont {Debs}}, \bibinfo {author}
  {\bibfnamefont {J.~D.}\ \bibnamefont {Close}}, \ and\ \bibinfo {author}
  {\bibfnamefont {N.~P.}\ \bibnamefont {Robins}},\ }\href
  {https://link.aps.org/doi/10.1103/PhysRevLett.117.138501} {\bibfield
  {journal} {\bibinfo  {journal} {Phys. Rev. Lett.}\ }\textbf {\bibinfo
  {volume} {117}},\ \bibinfo {pages} {138501} (\bibinfo {year}
  {2016})}\BibitemShut {NoStop}%
\bibitem [{\citenamefont {Kruse}\ \emph {et~al.}(2016)\citenamefont {Kruse},
  \citenamefont {Lange}, \citenamefont {Peise}, \citenamefont {L\"ucke},
  \citenamefont {Pezz\`e}, \citenamefont {Arlt}, \citenamefont {Ertmer},
  \citenamefont {Lisdat}, \citenamefont {Santos}, \citenamefont {Smerzi},\ and\
  \citenamefont {Klempt}}]{Kruse2016}%
  \BibitemOpen
  \bibfield  {author} {\bibinfo {author} {\bibfnamefont {I.}~\bibnamefont
  {Kruse}}, \bibinfo {author} {\bibfnamefont {K.}~\bibnamefont {Lange}},
  \bibinfo {author} {\bibfnamefont {J.}~\bibnamefont {Peise}}, \bibinfo
  {author} {\bibfnamefont {B.}~\bibnamefont {L\"ucke}}, \bibinfo {author}
  {\bibfnamefont {L.}~\bibnamefont {Pezz\`e}}, \bibinfo {author} {\bibfnamefont
  {J.}~\bibnamefont {Arlt}}, \bibinfo {author} {\bibfnamefont {W.}~\bibnamefont
  {Ertmer}}, \bibinfo {author} {\bibfnamefont {C.}~\bibnamefont {Lisdat}},
  \bibinfo {author} {\bibfnamefont {L.}~\bibnamefont {Santos}}, \bibinfo
  {author} {\bibfnamefont {A.}~\bibnamefont {Smerzi}}, \ and\ \bibinfo {author}
  {\bibfnamefont {C.}~\bibnamefont {Klempt}},\ }\href
  {https://link.aps.org/doi/10.1103/PhysRevLett.117.143004} {\bibfield
  {journal} {\bibinfo  {journal} {Phys. Rev. Lett.}\ }\textbf {\bibinfo
  {volume} {117}},\ \bibinfo {pages} {143004} (\bibinfo {year}
  {2016})}\BibitemShut {NoStop}%
\bibitem [{\citenamefont {Jachymski}\ \emph {et~al.}(2018)\citenamefont
  {Jachymski}, \citenamefont {Wasak}, \citenamefont {Idziaszek}, \citenamefont
  {Julienne}, \citenamefont {Negretti},\ and\ \citenamefont
  {Calarco}}]{Jachymski2018}%
  \BibitemOpen
  \bibfield  {author} {\bibinfo {author} {\bibfnamefont {K.}~\bibnamefont
  {Jachymski}}, \bibinfo {author} {\bibfnamefont {T.}~\bibnamefont {Wasak}},
  \bibinfo {author} {\bibfnamefont {Z.}~\bibnamefont {Idziaszek}}, \bibinfo
  {author} {\bibfnamefont {P.~S.}\ \bibnamefont {Julienne}}, \bibinfo {author}
  {\bibfnamefont {A.}~\bibnamefont {Negretti}}, \ and\ \bibinfo {author}
  {\bibfnamefont {T.}~\bibnamefont {Calarco}},\ }\href {\doibase
  10.1103/PhysRevLett.120.013401} {\bibfield  {journal} {\bibinfo  {journal}
  {Phys. Rev. Lett.}\ }\textbf {\bibinfo {volume} {120}},\ \bibinfo {pages}
  {013401} (\bibinfo {year} {2018})}\BibitemShut {NoStop}%
\bibitem [{\citenamefont {Meinert}\ \emph {et~al.}(2017)\citenamefont
  {Meinert}, \citenamefont {Knap}, \citenamefont {Kirilov}, \citenamefont
  {Jag-Lauber}, \citenamefont {Zvonarev}, \citenamefont {Demler},\ and\
  \citenamefont {N{\"a}gerl}}]{Meinert2017}%
  \BibitemOpen
  \bibfield  {author} {\bibinfo {author} {\bibfnamefont {F.}~\bibnamefont
  {Meinert}}, \bibinfo {author} {\bibfnamefont {M.}~\bibnamefont {Knap}},
  \bibinfo {author} {\bibfnamefont {E.}~\bibnamefont {Kirilov}}, \bibinfo
  {author} {\bibfnamefont {K.}~\bibnamefont {Jag-Lauber}}, \bibinfo {author}
  {\bibfnamefont {M.~B.}\ \bibnamefont {Zvonarev}}, \bibinfo {author}
  {\bibfnamefont {E.}~\bibnamefont {Demler}}, \ and\ \bibinfo {author}
  {\bibfnamefont {H.-C.}\ \bibnamefont {N{\"a}gerl}},\ }\href@noop {}
  {\bibfield  {journal} {\bibinfo  {journal} {Science}\ }\textbf {\bibinfo
  {volume} {356}},\ \bibinfo {pages} {945} (\bibinfo {year}
  {2017})}\BibitemShut {NoStop}%
\bibitem [{\citenamefont {Robens}\ \emph {et~al.}(2017)\citenamefont {Robens},
  \citenamefont {Zopes}, \citenamefont {Alt}, \citenamefont {Brakhane},
  \citenamefont {Meschede},\ and\ \citenamefont {Alberti}}]{Robens2017}%
  \BibitemOpen
  \bibfield  {author} {\bibinfo {author} {\bibfnamefont {C.}~\bibnamefont
  {Robens}}, \bibinfo {author} {\bibfnamefont {J.}~\bibnamefont {Zopes}},
  \bibinfo {author} {\bibfnamefont {W.}~\bibnamefont {Alt}}, \bibinfo {author}
  {\bibfnamefont {S.}~\bibnamefont {Brakhane}}, \bibinfo {author}
  {\bibfnamefont {D.}~\bibnamefont {Meschede}}, \ and\ \bibinfo {author}
  {\bibfnamefont {A.}~\bibnamefont {Alberti}},\ }\href {\doibase
  10.1103/PhysRevLett.118.065302} {\bibfield  {journal} {\bibinfo  {journal}
  {Phys. Rev. Lett.}\ }\textbf {\bibinfo {volume} {118}},\ \bibinfo {pages}
  {065302} (\bibinfo {year} {2017})}\BibitemShut {NoStop}%
\bibitem [{\citenamefont {Olshanii}(1998)}]{Olshanii1998}%
  \BibitemOpen
  \bibfield  {author} {\bibinfo {author} {\bibfnamefont {M.}~\bibnamefont
  {Olshanii}},\ }\href@noop {} {\bibfield  {journal} {\bibinfo  {journal}
  {Phys. Rev. Lett.}\ }\textbf {\bibinfo {volume} {81}},\ \bibinfo {pages}
  {938} (\bibinfo {year} {1998})}\BibitemShut {NoStop}%
\bibitem [{\citenamefont {Bergeman}\ \emph {et~al.}(2003)\citenamefont
  {Bergeman}, \citenamefont {Moore},\ and\ \citenamefont
  {Olshanii}}]{Bergeman2003}%
  \BibitemOpen
  \bibfield  {author} {\bibinfo {author} {\bibfnamefont {T.}~\bibnamefont
  {Bergeman}}, \bibinfo {author} {\bibfnamefont {M.}~\bibnamefont {Moore}}, \
  and\ \bibinfo {author} {\bibfnamefont {M.}~\bibnamefont {Olshanii}},\
  }\href@noop {} {\bibfield  {journal} {\bibinfo  {journal} {Phys. Rev. Lett.}\
  }\textbf {\bibinfo {volume} {91}},\ \bibinfo {pages} {163201} (\bibinfo
  {year} {2003})}\BibitemShut {NoStop}%
\bibitem [{\citenamefont {Granger}\ and\ \citenamefont
  {Blume}(2004)}]{Granger2004}%
  \BibitemOpen
  \bibfield  {author} {\bibinfo {author} {\bibfnamefont {B.~E.}\ \bibnamefont
  {Granger}}\ and\ \bibinfo {author} {\bibfnamefont {D.}~\bibnamefont
  {Blume}},\ }\href@noop {} {\bibfield  {journal} {\bibinfo  {journal} {Phys.
  Rev. Lett.}\ }\textbf {\bibinfo {volume} {92}},\ \bibinfo {pages} {133202}
  (\bibinfo {year} {2004})}\BibitemShut {NoStop}%
\bibitem [{\citenamefont {Kim}\ \emph {et~al.}(2006)\citenamefont {Kim},
  \citenamefont {Melezhik},\ and\ \citenamefont {Schmelcher}}]{Kim2006}%
  \BibitemOpen
  \bibfield  {author} {\bibinfo {author} {\bibfnamefont {J.~I.}\ \bibnamefont
  {Kim}}, \bibinfo {author} {\bibfnamefont {V.~S.}\ \bibnamefont {Melezhik}}, \
  and\ \bibinfo {author} {\bibfnamefont {P.}~\bibnamefont {Schmelcher}},\
  }\href {\doibase 10.1103/PhysRevLett.97.193203} {\bibfield  {journal}
  {\bibinfo  {journal} {Phys. Rev. Lett.}\ }\textbf {\bibinfo {volume} {97}},\
  \bibinfo {pages} {193203} (\bibinfo {year} {2006})}\BibitemShut {NoStop}%
\bibitem [{\citenamefont {Naidon}\ \emph {et~al.}(2007)\citenamefont {Naidon},
  \citenamefont {Tiesinga}, \citenamefont {Mitchell},\ and\ \citenamefont
  {Julienne}}]{Naidon2007}%
  \BibitemOpen
  \bibfield  {author} {\bibinfo {author} {\bibfnamefont {P.}~\bibnamefont
  {Naidon}}, \bibinfo {author} {\bibfnamefont {E.}~\bibnamefont {Tiesinga}},
  \bibinfo {author} {\bibfnamefont {W.~F.}\ \bibnamefont {Mitchell}}, \ and\
  \bibinfo {author} {\bibfnamefont {P.~S.}\ \bibnamefont {Julienne}},\
  }\href@noop {} {\bibfield  {journal} {\bibinfo  {journal} {New Journal of
  Physics}\ }\textbf {\bibinfo {volume} {9}},\ \bibinfo {pages} {19} (\bibinfo
  {year} {2007})}\BibitemShut {NoStop}%
\bibitem [{\citenamefont {Giannakeas}\ \emph {et~al.}(2012)\citenamefont
  {Giannakeas}, \citenamefont {Diakonos},\ and\ \citenamefont
  {Schmelcher}}]{Giannakeas2012}%
  \BibitemOpen
  \bibfield  {author} {\bibinfo {author} {\bibfnamefont {P.}~\bibnamefont
  {Giannakeas}}, \bibinfo {author} {\bibfnamefont {F.~K.}\ \bibnamefont
  {Diakonos}}, \ and\ \bibinfo {author} {\bibfnamefont {P.}~\bibnamefont
  {Schmelcher}},\ }\href {\doibase 10.1103/PhysRevA.86.042703} {\bibfield
  {journal} {\bibinfo  {journal} {Phys. Rev. A}\ }\textbf {\bibinfo {volume}
  {86}},\ \bibinfo {pages} {042703} (\bibinfo {year} {2012})}\BibitemShut
  {NoStop}%
\bibitem [{\citenamefont {Giannakeas}\ \emph {et~al.}(2013)\citenamefont
  {Giannakeas}, \citenamefont {Melezhik},\ and\ \citenamefont
  {Schmelcher}}]{Giannakeas2013}%
  \BibitemOpen
  \bibfield  {author} {\bibinfo {author} {\bibfnamefont {P.}~\bibnamefont
  {Giannakeas}}, \bibinfo {author} {\bibfnamefont {V.~S.}\ \bibnamefont
  {Melezhik}}, \ and\ \bibinfo {author} {\bibfnamefont {P.}~\bibnamefont
  {Schmelcher}},\ }\href@noop {} {\bibfield  {journal} {\bibinfo  {journal}
  {Phys. Rev. Lett.}\ }\textbf {\bibinfo {volume} {111}},\ \bibinfo {pages}
  {183201} (\bibinfo {year} {2013})}\BibitemShut {NoStop}%
\bibitem [{\citenamefont {Peng}\ \emph {et~al.}(2014)\citenamefont {Peng},
  \citenamefont {Tan},\ and\ \citenamefont {Jiang}}]{Peng2014}%
  \BibitemOpen
  \bibfield  {author} {\bibinfo {author} {\bibfnamefont {S.-G.}\ \bibnamefont
  {Peng}}, \bibinfo {author} {\bibfnamefont {S.}~\bibnamefont {Tan}}, \ and\
  \bibinfo {author} {\bibfnamefont {K.}~\bibnamefont {Jiang}},\ }\href@noop {}
  {\bibfield  {journal} {\bibinfo  {journal} {Phys. Rev. Lett.}\ }\textbf
  {\bibinfo {volume} {112}},\ \bibinfo {pages} {250401} (\bibinfo {year}
  {2014})}\BibitemShut {NoStop}%
\bibitem [{\citenamefont {He{\ss}}\ \emph {et~al.}(2014)\citenamefont
  {He{\ss}}, \citenamefont {Giannakeas},\ and\ \citenamefont
  {Schmelcher}}]{Hess2014}%
  \BibitemOpen
  \bibfield  {author} {\bibinfo {author} {\bibfnamefont {B.}~\bibnamefont
  {He{\ss}}}, \bibinfo {author} {\bibfnamefont {P.}~\bibnamefont {Giannakeas}},
  \ and\ \bibinfo {author} {\bibfnamefont {P.}~\bibnamefont {Schmelcher}},\
  }\href@noop {} {\bibfield  {journal} {\bibinfo  {journal} {Phys. Rev. A}\
  }\textbf {\bibinfo {volume} {89}},\ \bibinfo {pages} {052716} (\bibinfo
  {year} {2014})}\BibitemShut {NoStop}%
\bibitem [{\citenamefont {Idziaszek}\ \emph {et~al.}(2015)\citenamefont
  {Idziaszek}, \citenamefont {Jachymski},\ and\ \citenamefont
  {Julienne}}]{Idziaszek2015}%
  \BibitemOpen
  \bibfield  {author} {\bibinfo {author} {\bibfnamefont {Z.}~\bibnamefont
  {Idziaszek}}, \bibinfo {author} {\bibfnamefont {K.}~\bibnamefont
  {Jachymski}}, \ and\ \bibinfo {author} {\bibfnamefont {P.~S.}\ \bibnamefont
  {Julienne}},\ }\href@noop {} {\bibfield  {journal} {\bibinfo  {journal} {New
  Journal of Physics}\ }\textbf {\bibinfo {volume} {17}},\ \bibinfo {pages}
  {035007} (\bibinfo {year} {2015})}\BibitemShut {NoStop}%
\bibitem [{\citenamefont {He{\ss}}\ \emph {et~al.}(2015)\citenamefont
  {He{\ss}}, \citenamefont {Giannakeas},\ and\ \citenamefont
  {Schmelcher}}]{Hess2015}%
  \BibitemOpen
  \bibfield  {author} {\bibinfo {author} {\bibfnamefont {B.}~\bibnamefont
  {He{\ss}}}, \bibinfo {author} {\bibfnamefont {P.}~\bibnamefont {Giannakeas}},
  \ and\ \bibinfo {author} {\bibfnamefont {P.}~\bibnamefont {Schmelcher}},\
  }\href@noop {} {\bibfield  {journal} {\bibinfo  {journal} {Phys. Rev. A}\
  }\textbf {\bibinfo {volume} {92}},\ \bibinfo {pages} {022706} (\bibinfo
  {year} {2015})}\BibitemShut {NoStop}%
\bibitem [{\citenamefont {Melezhik}\ and\ \citenamefont
  {Negretti}(2016)}]{Melezhik2016}%
  \BibitemOpen
  \bibfield  {author} {\bibinfo {author} {\bibfnamefont {V.~S.}\ \bibnamefont
  {Melezhik}}\ and\ \bibinfo {author} {\bibfnamefont {A.}~\bibnamefont
  {Negretti}},\ }\href@noop {} {\bibfield  {journal} {\bibinfo  {journal}
  {Phys. Rev. A}\ }\textbf {\bibinfo {volume} {94}},\ \bibinfo {pages} {022704}
  (\bibinfo {year} {2016})}\BibitemShut {NoStop}%
\bibitem [{\citenamefont {Jachymski}\ \emph {et~al.}(2017)\citenamefont
  {Jachymski}, \citenamefont {Meinert}, \citenamefont {Veksler}, \citenamefont
  {Julienne},\ and\ \citenamefont {Fishman}}]{Jachymski2017}%
  \BibitemOpen
  \bibfield  {author} {\bibinfo {author} {\bibfnamefont {K.}~\bibnamefont
  {Jachymski}}, \bibinfo {author} {\bibfnamefont {F.}~\bibnamefont {Meinert}},
  \bibinfo {author} {\bibfnamefont {H.}~\bibnamefont {Veksler}}, \bibinfo
  {author} {\bibfnamefont {P.~S.}\ \bibnamefont {Julienne}}, \ and\ \bibinfo
  {author} {\bibfnamefont {S.}~\bibnamefont {Fishman}},\ }\href {\doibase
  10.1103/PhysRevA.95.052703} {\bibfield  {journal} {\bibinfo  {journal} {Phys.
  Rev. A}\ }\textbf {\bibinfo {volume} {95}},\ \bibinfo {pages} {052703}
  (\bibinfo {year} {2017})}\BibitemShut {NoStop}%
\bibitem [{\citenamefont {Chin}\ \emph {et~al.}(2010)\citenamefont {Chin},
  \citenamefont {Grimm}, \citenamefont {Julienne},\ and\ \citenamefont
  {Tiesinga}}]{Chin2010}%
  \BibitemOpen
  \bibfield  {author} {\bibinfo {author} {\bibfnamefont {C.}~\bibnamefont
  {Chin}}, \bibinfo {author} {\bibfnamefont {R.}~\bibnamefont {Grimm}},
  \bibinfo {author} {\bibfnamefont {P.}~\bibnamefont {Julienne}}, \ and\
  \bibinfo {author} {\bibfnamefont {E.}~\bibnamefont {Tiesinga}},\ }\href@noop
  {} {\bibfield  {journal} {\bibinfo  {journal} {Rev. Mod. Phys.}\ }\textbf
  {\bibinfo {volume} {82}},\ \bibinfo {pages} {1225} (\bibinfo {year}
  {2010})}\BibitemShut {NoStop}%
\bibitem [{\citenamefont {Gribakin}\ and\ \citenamefont
  {Flambaum}(1993)}]{Gribakin}%
  \BibitemOpen
  \bibfield  {author} {\bibinfo {author} {\bibfnamefont {G.~F.}\ \bibnamefont
  {Gribakin}}\ and\ \bibinfo {author} {\bibfnamefont {V.~V.}\ \bibnamefont
  {Flambaum}},\ }\href@noop {} {\bibfield  {journal} {\bibinfo  {journal}
  {Phys. Rev. A}\ }\textbf {\bibinfo {volume} {48}},\ \bibinfo {pages} {546}
  (\bibinfo {year} {1993})}\BibitemShut {NoStop}%
\bibitem [{\citenamefont {Gao}(1998)}]{Gao1998}%
  \BibitemOpen
  \bibfield  {author} {\bibinfo {author} {\bibfnamefont {B.}~\bibnamefont
  {Gao}},\ }\href@noop {} {\bibfield  {journal} {\bibinfo  {journal} {Phys.
  Rev. A}\ }\textbf {\bibinfo {volume} {58}},\ \bibinfo {pages} {1728}
  (\bibinfo {year} {1998})}\BibitemShut {NoStop}%
\bibitem [{\citenamefont {Gao}(2000)}]{Gao2000}%
  \BibitemOpen
  \bibfield  {author} {\bibinfo {author} {\bibfnamefont {B.}~\bibnamefont
  {Gao}},\ }\href@noop {} {\bibfield  {journal} {\bibinfo  {journal} {Phys.
  Rev. A}\ }\textbf {\bibinfo {volume} {62}},\ \bibinfo {pages} {050702}
  (\bibinfo {year} {2000})}\BibitemShut {NoStop}%
\bibitem [{\citenamefont {Massignan}\ and\ \citenamefont
  {Castin}(2006)}]{Massignan2006}%
  \BibitemOpen
  \bibfield  {author} {\bibinfo {author} {\bibfnamefont {P.}~\bibnamefont
  {Massignan}}\ and\ \bibinfo {author} {\bibfnamefont {Y.}~\bibnamefont
  {Castin}},\ }\href@noop {} {\bibfield  {journal} {\bibinfo  {journal} {Phys.
  Rev. A}\ }\textbf {\bibinfo {volume} {74}},\ \bibinfo {pages} {013616}
  (\bibinfo {year} {2006})}\BibitemShut {NoStop}%
\bibitem [{\citenamefont {Braunstein}\ and\ \citenamefont
  {Caves}(1994)}]{braunstein1994statistical}%
  \BibitemOpen
  \bibfield  {author} {\bibinfo {author} {\bibfnamefont {S.~L.}\ \bibnamefont
  {Braunstein}}\ and\ \bibinfo {author} {\bibfnamefont {C.~M.}\ \bibnamefont
  {Caves}},\ }\href {\doibase 10.1103/PhysRevLett.72.3439} {\bibfield
  {journal} {\bibinfo  {journal} {Phys. Rev. Lett.}\ }\textbf {\bibinfo
  {volume} {72}},\ \bibinfo {pages} {3439} (\bibinfo {year}
  {1994})}\BibitemShut {NoStop}%
\bibitem [{\citenamefont {R{\'e}fr{\'e}gier}(2012)}]{refregier2012noise}%
  \BibitemOpen
  \bibfield  {author} {\bibinfo {author} {\bibfnamefont {P.}~\bibnamefont
  {R{\'e}fr{\'e}gier}},\ }\href@noop {} {\emph {\bibinfo {title} {Noise theory
  and application to physics: from fluctuations to information}}}\ (\bibinfo
  {publisher} {Springer Science \& Business Media},\ \bibinfo {year}
  {2012})\BibitemShut {NoStop}%
\bibitem [{\citenamefont {Cram{\'e}r}(2016)}]{cramer2016mathematical}%
  \BibitemOpen
  \bibfield  {author} {\bibinfo {author} {\bibfnamefont {H.}~\bibnamefont
  {Cram{\'e}r}},\ }\href@noop {} {\emph {\bibinfo {title} {Mathematical Methods
  of Statistics (PMS-9)}}},\ Vol.~\bibinfo {volume} {9}\ (\bibinfo  {publisher}
  {Princeton university press},\ \bibinfo {year} {2016})\BibitemShut {NoStop}%
\bibitem [{\citenamefont {Gross}\ and\ \citenamefont
  {Bloch}(2017)}]{Gross2017}%
  \BibitemOpen
  \bibfield  {author} {\bibinfo {author} {\bibfnamefont {C.}~\bibnamefont
  {Gross}}\ and\ \bibinfo {author} {\bibfnamefont {I.}~\bibnamefont {Bloch}},\
  }\href@noop {} {\bibfield  {journal} {\bibinfo  {journal} {Science}\ }\textbf
  {\bibinfo {volume} {357}},\ \bibinfo {pages} {996} (\bibinfo {year}
  {2017})}\BibitemShut {NoStop}%
\bibitem [{\citenamefont {Helstrom}(1976)}]{helstrom1976quantum}%
  \BibitemOpen
  \bibfield  {author} {\bibinfo {author} {\bibfnamefont {C.~W.}\ \bibnamefont
  {Helstrom}},\ }\href@noop {} {\emph {\bibinfo {title} {Quantum detection and
  estimation theory}}}\ (\bibinfo  {publisher} {Academic Press},\ \bibinfo
  {year} {1976})\BibitemShut {NoStop}%
\end{thebibliography}%
\end{document}